\documentclass[12pt]{article}
\pdfoutput=1
\usepackage{float}
\usepackage[final]{graphicx}
\usepackage{amssymb,amsmath}
\usepackage{commath}
\usepackage{epsfig}
\usepackage{epstopdf}
\usepackage{latexsym}
\usepackage{graphicx}
\usepackage{subfigure}
\usepackage{caption}
\usepackage{booktabs}
\usepackage{tkz-euclide}

\usepackage{makeidx}
\usepackage{cite}
\usepackage{bm}
\usepackage{geometry}
\usepackage{braket}
\usepackage[toc]{appendix}

\usepackage{tikz}
\usepackage{xspace}

\usepackage{xcolor}

\usetikzlibrary{calc,decorations.markings,arrows.meta,shapes.misc,decorations.pathmorphing,calc,bending}

\usetikzlibrary{arrows.meta,shapes.misc,decorations.pathmorphing,calc,bending}

\tikzset{
    branch point/.style={cross out,draw=black,fill=none,minimum size=2*(#1-\pgflinewidth),inner sep=0pt,outer sep=0pt},
    branch point/.default=5
}
\tikzset{
    branch cut/.style={
        decorate,decoration=snake,
        to path={
            (\tikztostart) -- (\tikztotarget) \tikztonodes
        },
    }
}

\usepackage{color}
\usepackage{datetime}

\ifpdf
\fi

\def\cM{{\cal M}}

\def\cO{{\cal O}}

\def\cL{{\cal L}}

\def\cL{{\cal L}}

\def\cC{{\cal C}}

\def\cM{{\cal M}}

\definecolor{cardinal}{rgb}{0.6,0,0}
\definecolor{darkgreen}{rgb}{0,0.5,0}
\definecolor{golden}{rgb}{0.92, 0.7, 0}
\definecolor{midnight}{rgb}{0, 0, 0.5}
\definecolor{darkblue}{rgb}{0.2, 0, 0.8}

\newcommand{\be}{\begin{equation}}
    \newcommand{\ee}{\end{equation}}
\newcommand{\bea}{\begin{eqnarray}}
    \newcommand{\eea}{\end{eqnarray}}

\newcommand{\Poincare}{Poincar\'e\xspace}

\begin{document}

    \begin{titlepage}

        \bigskip        \bigskip        \bigskip
        \centerline{\large \bf Brane Detectors of a Dynamical Phase Transition in a Driven CFT}
\bigskip
\bigskip
\centerline{\bf Suchetan~Das$^{1}$, Bobby~Ezhuthachan$^2$,
Arnab~Kundu$^{3,4}$, } \centerline{\bf Somnath~Porey$^2$,
Baishali~Roy$^2$, K.~Sengupta$^5$}
\bigskip
\centerline{$^1$Department of Physics, Indian Institute of Technology Kanpur, Kanpur 208016, India.}
\bigskip
\centerline{$^2$Ramakrishna Mission Vivekananda Educational and Research Institute,}
\centerline{Belur Math, Howrah-711202, West Bengal, India.}
\bigskip
\centerline{$^3$ Saha Institute of Nuclear Physics, 1/AF, Bidhannagar, Kolkata 700064, India.}
\centerline{$^4$Homi Bhaba National Institute, Training School Complex, Anushaktinagar, Mumbai 400094, India.}
\bigskip
\centerline{$^5$ School of Physical Sciences, Indian Association for the Cultivation of Science,}
\centerline{2A and 2B Raja S.C.Mullick Road, Jadavpur, Kolkata-700032, West Bengal, India.}
\bigskip
\centerline{suchetan[at]iitk.ac.in,  bobby.phy[at]gm.rkmvu.ac.in, arnab.kundu[at]saha.ac.in}
\centerline{somnathhimu00[at]gm.rkmvu.ac.in, baishali.roy025[at]gm.rkmvu.ac.in, tpks[at]iacs.res.in }

\begin{abstract}

\noindent We show that a dynamical transition from a non-heating
to a heating phase of a periodic $SL(2,\mathbb{R})$ driven two dimensional conformal field
theory (CFT) with a large central charge is perceived as a first
order transition by a bulk brane embedded in the dual AdS. We
construct the dual bulk metric corresponding to a driven CFT for
both the heating and the non-heating phases. These metrics are different AdS$_{2}$ slices of the pure AdS$_{3}$ metric. We embed a brane in the
obtained dual AdS space and provide an explicit computation of its free
energy both in the probe limit and for an end-of-world (EOW) brane
taking into account its backreaction. Our analysis indicates a
finite discontinuity in the first derivative of the brane free
energy as one moves from the non-heating to the heating phase (by
tuning the drive amplitude and/or frequency of the driven CFT) thus
demonstrating the presence of the bulk first order transition.
Interestingly, no such transition is perceived by the bulk in the
absence of the brane. We also provide explicit computations of two-point, four-point out-of-time correlators (OTOC) using the bulk
picture. Our analysis shows that the structure of these correlators in different phases match their counterparts computed in the driven CFT. We
analyze the effect of multiple EOW branes in the bulk and discuss
possible extensions of our work for richer geometries and branes.

\end{abstract}

        \newpage


    \end{titlepage}
    \tableofcontents


    \rule{\textwidth}{.5pt}\\

\numberwithin{equation}{section}
    \section{Introduction}

Non-equilibrium dynamics of driven quantum matter has been
extensively studied in the recent
past\cite{rev1,rev2,rev3,rev4,rev5,rev6,rev7,rev8,rev9,exp1,exp2,exp3,
exp4}. Out of the several drive protocols that can be used to take a
system out of equilibrium,  periodic drives, whose stroboscopic
dynamics can be described by Floquet Hamiltonians, have received the
most attention \cite{rev5,rev8,magrev,fpt1,fpt2}. The reason for
this stems from several phenomena such as dynamical freezing
\cite{df1,df2,df3,df4,df5}, dynamical localization
\cite{dloc1,dloc2,dloc3,dloc4}, topological transitions in driven
systems \cite{topo1,topo2,topo3,topo4}, realization of time
crystalline states \cite{tc1,tc2,tc3,tc4}, dynamical transitions
\cite{dtran1,dtran2,dtran3,dtran4}, and tuning ergodicity properties
of quantum systems \cite{erg1,erg2}; these phenomena have no
analogue in either equilibrium or aperiodically driven
non-equilibrium systems.

Several recent studies have focussed on the effect of both quench \cite
{Wen:2018vux} and periodic drives \cite{Wen:2018agb, Wen:2020wee,
Han:2020kwp, Fan:2020orx, Andersen:2020xvu, Das:2021gts} on
conformal field theories. The studies involving periodic protocols
usually consider a Hamiltonian which is expressed in terms of
standard Virasoro generators $L_0$ and $L_{\pm 1}$ and is therefore
valued in an $su(1,1)$-algebra. The periodic drive in such models
leads to an evolution operator $U$ which is valued in ${\rm
SU}(1,1)$. It is well-known that such a dynamics leads to two
distinct phases separated by a dynamical transition. These are the
heating (hyperbolic) and the non-heating (elliptic) phases; the
Casimir of the $su(1,1)$ algebra has opposite signs in the two
phases. The transition line where the Casimir vanishes is often
referred to as the parabolic line. The presence of a periodic drive,
characterized by a frequency $\omega_D= 2\pi/T$, where $T$ is the
time period, allows one to access this dynamic transition by tuning
the drive frequency. Equivalently, such a tuning is possible by
changing the drive amplitude.

The AdS/CFT correspondence stipulates that corresponding to every
such $(1+1)$D CFT, there exists a three-dimensional ($3$D) dual AdS
bulk \cite{ads1}. \footnote{Note that the CFT dual of pure $3$D
gravity is a debated issue. A precise duality generally contains
various fluxes in a $10$D bulk geometry. However, we will ignore
this issue for now, as we will be focussing on a rather generic
point which is expected to remain qualitatively true.} In fact,
using this correspondence, one can have a definite procedure for
extending the CFT Hamiltonian in the bulk \cite{Anand:2017dav}. It
is then natural to ask what constitutes the bulk signature of the
dynamic transition of a driven CFT. This is the central question
which we aim to study in this work. More precisely, the main goal of
our work is to geometrize the dynamical phase transition and to
provide a precise and explicit $3$D geometric and Holographic
construction that captures this transition.

The main points of our work can be summarized as follows. First, we
show that there are two key steps to construct the geometric
description mentioned above. We take CFT$_{2}$ vacuum as the reference state which is dual to pure AdS$_{3}$. Since the vacuum does not evolve under $sl(2,R)$ valued Floquet-Hamiltonian, the bulk geometry remains pure AdS. However to geometrize the drive, the basic ingredient is the set of bulk
generators corresponding to the Virasoro generators $L_0$ and
$L_{\pm 1}$ of the boundary CFT\cite{Anand:2017dav}. Subsequently,
one finds the curves in the bulk geometry which are generated by the
bulk Hamiltonian corresponding to the the  CFT Floquet-Hamiltonian.
These curves inherit a natural induced metric on them, which are
simply patches of the AdS$_3$-spacetime. These patches have a
natural AdS$_2$ slicing, which differ for each phase. In particular,
for the heating phase, we find a AdS$_2$ black hole slice. For the
non-heating phase we find a global AdS$_2$ slicing and the phase
boundary corresponds to a Poincare AdS$_2$ slicing. While these
patches are highly suggestive, the heating and the non-heating
phases cannot be distinguished by the corresponding free energy,
which is given by the Euclidean on-shell action for pure $3$D
gravity in AdS$_3$. At this point, it is worth emphasizing that the
boundary CFT Hamiltonian is $sl(2,R)$-valued and therefore there are
no large gauge transformations in the bulk. Thus, the heating and
the non-heating phases are identical in terms of their Euclidean
on-shell actions.

Second, we find that the crucial ingredient in distinguishing
between these CFT phases is a brane degree of freedom. These branes
are co-dimension one hypersurfaces in the bulk geometry. In this
work, we have considered both probe branes as well as end-of-world
(EOW) branes. These are respectively probes and fully back-reacting
objects in the AdS$_3$ geometry. Given a particular patch (corresponding to the heating, the non-heating or the phase boundary), these
branes can distinguish between the on-shell Euclidean action of the
(gravity + brane)-system. The central result of this work is that
the corresponding free energy displays a first order phase
transition of the combined system, which is a close cousin of the
Hawking-Page transition\cite{Hawking:1982dh}. We explicitly demonstrate that this first
order transition also occurs due to the change in sign of the
Casimir of the boundary CFT, thereby establishing a direct link
between it and the dynamical transition of the driven CFT mentioned
above.

Third, we generalize this construction and introduce more than one
EOW-branes. As an example, we have considered two EOW-branes, which
results in a rich structure associated to the phase transition. The
basic qualitative features of the phase transition remain the same.
We note that insertion of the EOW-branes correspond to inserting
conformal boundaries to the boundary CFT. Correspondingly, the CFT
is defined on a strip with an infinite family of boundary
conditions. These boundary conditions are labelled by the respective
brane tensions valued in the range of $[0,1]$. Especially, this
family of boundary conditions naturally allow for exciting a
boundary-condition-changing operator in the CFT, whenever the
boundary conditions are non-identical at the end points.

Finally, we also study the signature of different phases in unequal
time two point and higher point correlators under the $sl(2,R)$ drive
\cite{Das:2021gts}, \cite{Das:2022jrr1} from the bulk gravity
picture without inserting any brane. In this context, we describe
how the different AdS$_2$ slicing corresponding to the two phases
and the phase boundary are crucial in determining different temporal
behavior of $2$-point and $4$-point functions from the Holographic
description, which can be matched with results available from large
$c$ CFT computations. In particular, we match the two point
functions in each phase with a direct boundary computation of the
two point function. We then set-up the OTOC computation in the bulk.
This involves, the by now well-known method of, computing a two
point function of an operator in a shock wave geometry created by
the other operator\cite{Roberts:2014ifa1, Shenker:2013pqa1}. The
AdS$_2$ black hole slicing for the heating phase is crucial to
obtain the exponential temporal growth of the OTOC \footnote{We want
to emphasize again that the dual 3d AdS metric has no horizon and
the blackhole resides in it's AdS$_{2}$ slice. To the best of our
knowledge, OTOCs in this kind of geometries has not been studied
earlier.}, that we had obtained in our previous
work\cite{Das:2022jrr1}. We show that the Lyapunov exponent matches
exactly with the boundary computation. Moreover we also show how the
other AdS$_2$ slicing corresponding to the non heating phase and the
phase boundary results in an oscillatory and power law temporal
growth of the OTOCs in these two phases.

The plan for the rest of the paper is as follows. In Sec.\
\ref{bd1}, we construct the bulk metric for the different phases of
the driven CFTs. This is followed by Sec.\ \ref{bi1}, where we
discuss embedding both probe and EOW branes. Next, in Sec.\
\ref{corr1}, we compute two-point correlation functions and four-point OTOC
from the bulk in the large $c$ limit and compare these results  with the
corresponding counterparts obtained from the driven CFT at the
boundary. Finally, we discuss our main results and the possibilities
of their further extension and conclude in Sec.\ \ref{diss}.

\numberwithin{equation}{section}
\section{Bulk metrics in different phases subjected to an $SL(2,\mathbb{R})$ drive}
\label{bd1}

In this section, we construct the bulk metrics for the various
phases of the $SL(2,\mathbb{R})$ driven CFT.

\paragraph{General strategy:}

To begin with let us consider a generic state $|\psi\rangle$ that is
evolved in stroboscopic time $n$ under some $2$D boundary
periodically driven Hamiltonian $H$. We want to understand the three
dimensional holographic realization of the state as well as it's
evolution i.e.
\begin{align}
|\psi(n)\rangle = U(nT,0) |\psi\rangle = e^{-i H_F n T/\hbar} |\psi\rangle \ ,
\end{align}
where $U(nT,0)$ is the evolution operator and $H_F$ is the Floquet
Hamiltonian. Here $n$ is  a positive integer, $T=2\pi/\omega_D$ is
the drive period, and $\omega_D$ is the drive frequency. The
complete holographic picture could be obtained by a two step
process:
\begin{itemize}
\item First we find the geometric dual of a one parameter class of states of the above form, with $nT$ replaced by $s$. At this step, $s$ (or rather $s/T$) should be interpreted as a real parameter of the geometric description. The geometric dual can then be found by solving the Einstein equation with source given by the expectation value of boundary stress tensor in certain choice of coordinate system. 

\item The next step is to rewrite the new metric in a parametrization where $s$ itself becomes the time in the metric. This means going to the frame of the co-moving observer along the  curve generated by the Floquet Hamiltonian itself. The well-known example is the Rindler wedge which is obtained by changing the flat spacetime coordinates to new coordinates generated by the trajectory of an accelerated observer. Once we are able to get the effective boundary metric by parametrizing the boundary curve, we need to lift that into the bulk \cite{Skenderis:1999nb, MacCormack:2018rwq, Goto:2021sqx, Lapierre:2019rwj}. One straightforward yet harder way to get the final bulk metric is to solve the same Einstein equation as in the first step with boundary metric as the boundary condition. However, our task is simpler: We can directly solve for the bulk curves generated by the bulk representation of the boundary Floquet Hamiltonian. The curves will be parametrized by $s$ and 
other intrinsic co-ordinates, in terms of which one rewrites the above metric.

\end{itemize}

The simplest example of the above set up is when we take $H=H_{\rm CFT}=L_{0}+\bar{L}_{0}$ and the state is the vacuum $|0\rangle$. The state does not evolve in (the stroboscopic) time as $e^{i(L_{0}+\bar{L}_{0})n}|0\rangle = |0\rangle$. In the Euclidean boundary, this corresponds to radial quantization which can be visualized by conformally mapping the plane into a cylinder, where $n$ coincides with the time direction in the cylinder. On the bulk, this corresponds to global AdS$_{3}$ with $n$ naturally enlarged to $s$ which acts as the global time. If we choose $H$ to be other linear combinations of conformal generators $L_{p},\bar{L}_{p}$, it corresponds to a different quantization \cite{Ishibashi:2015jba},\cite{Ishibashi:2016bey} in the CFT. The corresponding bulk metric will be obtained by mapping it from the AdS$_{3}$ coordinates under large diffeomorphism (generated by boundary $L_{p}$'s) in a specific gauge \cite{Caputa:2022zsr} and then solving for the bulk curve. Let us now discuss this explicitly when $p=\{0,\pm 1\}$.

\paragraph{Bulk metric under an $SL(2,\mathbb{R})$ drive:}

To compute the bulk metric in an $SL(2,\mathbb{R})$ driven CFT, we
extend the boundary Hamiltonian into the bulk by replacing the
global Virasoro generators by it's AdS$_{3}$ representation
\cite{Anand:2017dav}:
    \begin{equation}\label{a1}
        L_{b,0}=-\frac{1}{2}z\partial_{z}-\zeta\partial_{\zeta}\ , L_{b,1}=\frac{1}{2}z\zeta\partial_{z}+\zeta^{2}\partial_{\zeta}-z^{2}\partial_{\bar{\zeta}} \ , L_{b,-1}=\partial_{\zeta} \ .
    \end{equation}
Here, $\zeta=x-i\tau$ and $\bar{\zeta}=x+i\tau$ are the boundary coordinates and $z$ is along the bulk direction. We will work in two
step discrete drive protocol \cite{Wen:2018agb, Wen:2020wee,
Han:2020kwp} governed by the Hamiltonian
$H_{\phi}=L_{0}-\frac{1}{2}\tanh(2\phi)(L_{1}+L_{-1})+\text{anti
chiral part}$. For time period $\tau_{0}$ the system is evolved by
$H_{0}=H_{\phi=0}\ $, and for time period $\tau_{1}$ it is evolved
by $H_{1}=H_{\phi\neq 0}$ and then we repeat it periodically for $n$
number of drives. The stroboscopic time parameter $n$ plays the role
of time in our setting. Here we still start with the vacuum
$|0\rangle$. Since again $H_{\phi}$ is constructed out of
$SL(2,\mathbb{R})$ generators, the vacuum remains unchanged. This
dictates that the bulk remains pure AdS$_{3}$. However to construct
the bulk tangent curve along the direction of drive $n$, it might be
difficult to track down the time dependent set up at each period of
time. For computational purposes, it would be useful to find the
Floquet Hamiltonian which controls the evolution of the driven
system after an integer number of drive periods. We can write an
effective Hamiltonian with the following form
\begin{align}
H_{\rm eff}=\alpha(L_{0}+\bar{L}_{0})+\beta(L_{1}+\bar{L}_{1})+\gamma(L_{-1}+\bar{L}_{-1}) \ .
\end{align}
We give the relevant details of $\alpha,\beta,\gamma$ in the appendix \ref{B}.
 The corresponding AdS$_{3}$ representation of the Hamiltonian is given by,
    \begin{equation}\label{a2}
        H_{b}=\alpha(L_{b,0}+\bar{L}_{b,0})+\beta(L_{b,1}+\bar{L}_{b,1})+\gamma(L_{b,-1}+\bar{L}_{b,-1}) \ ,
    \end{equation}
     This class of Hamiltonians are valued in $su(1,1)$-algebra which generates a time-evolution valued in the $SU(1,1)$ group. Given $\{\alpha, \beta, \gamma\}$, the Casimir of the algebra is given by $(\alpha^2 - 4 \beta \gamma)$ and we define:
\begin{eqnarray}
d = \frac{\alpha^2 - 4 \beta \gamma}{4\beta^2}  \ ,
\end{eqnarray}
which keeps track of the sign of the Casimir. It is now well-known that the system can be tuned to any of the three distinct phases, depending on the sign of $d$, see {\it e.g.}~\cite{Wen:2018agb, Wen:2020wee, Han:2020kwp, Fan:2020orx, Andersen:2020xvu, Das:2021gts} for several related works exploring these phases.

Specifically, we can distinguish the three phases of the system:
    \begin{align*}
        & d<0 : \text{Heating Phase} \ , \\
        & d>0 : \text{Non-heating Phase}\ , \\
        & d=0 : \text{Phase transition} \ .
    \end{align*}
Substituting \eqref{a1} and corresponding complex conjugates in \eqref{a2} we can write:
    \begin{equation}
        H_b=(-\alpha z+2 \beta z x)\partial_z+(-\alpha x-\beta z^2+\beta(x^2-\tau^2)+\gamma)\partial_x+(-\alpha \tau+2 \beta x \tau)\partial_\tau \ .
    \end{equation}
As mentioned before, we consider an intrinsic coordinate, denoted by $s\in {\mathbb R}$, to parameterize the curves generated by the bulk Hamiltonian and solve the following tangent equations\cite{Das:2019iit}:
    \begin{align}
        \frac{dz(s)}{ds} & = -\alpha z+2 \beta z x \ , \label{em1}\\
        \frac{d \tau(s)}{ds} & = -\alpha \tau+2 \beta x \tau \ , \label{em2}\\
        \frac{dx(s)}{ds}& = -\alpha x-\beta z^2+\beta(x^2-\tau^2)+\gamma \ , \label{em3}
    \end{align} to find the relation between the embedding coordinates $\{\tau, x, z\}$ and the patch solved by the equations in (\ref{em1})-(\ref{em3}). Note that, the bulk coordinate $s$ is continuous while the stroboscopic time is discrete. The identification of the stroboscopic time with this continuous bulk coordinate is made only at discrete points. Said another way, different values of the stroboscopic time correspond to different points on the curve whose coordinate is $s$. Note also, that a solution to the above equations will allow an arbitrary constant shift in $s$ and therefore, effectively we can set the range: $s \in [-\infty, \infty]$. The solution space can be divided into three categories, depending on the sign of $d$. Below, we discuss these in detail.

    \vskip 0.5cm
\noindent   \textbf{\underline{For non-heating phase(d$>0$)}}:

When $d>0$, the set of equations in (\ref{em1})-(\ref{em3}) is solved by
    \begin{align}\label{a3}
        x & = -\frac{\sqrt{d}}{2}\Bigl(\coth[\mu(s+i\theta)]+\coth[\mu(s-i\theta)]\Bigl)  \ , \nonumber\\
        \tau & =-\frac{\sqrt{d}}{2 i\sqrt{1+c_1^2}}\Bigl(\coth[\mu(s+i\theta)]-\coth[\mu(s-i\theta)]\Bigl)  \ , \\
        z & =-\frac{\sqrt{d} \ c_1}{i\sqrt{1+c_1^2}}\Bigl(\coth[\mu(s+i\theta)]-\coth[\mu(s-i\theta)]\Bigl)  \ , \quad \mu=\beta\sqrt{d} \ .
    \end{align}
    Here $c_1$ and $\theta$ characterize the parametric solutions.  After rewriting $c_1 = \tan \phi_1$ and substituting \eqref{a3} in AdS$_3$-\Poincare metric, $ds^2=\frac{dx^2+d\tau^2+dz^2}{z^2}$ we get\footnote{When we are substituting in the metric, we treat $\theta$ and $\phi$ to be the normal coordinates to the curve.}
    \begin{equation}\label{me1}
        ds^2=\frac{d\phi^2}{{\sin}^2[\phi]}+\frac{4 \beta^2 d\  (ds^2+d\theta^2)}{{\sin}^2[\phi]\ {\sin}^2[2\sqrt{d}\beta\theta]} \ .
    \end{equation}
Finally, analytically continuing $s\rightarrow is$ we obtain:
    \begin{equation}\label{m1}
        ds^2=\frac{d\phi^2}{{\sin}^2[\phi]}+\frac{4 \beta^2 d\  (-ds^2+d\theta^2)}{{\sin}^2[\phi]\ {\sin}^2[2\sqrt{d}\beta\theta]} \ .
    \end{equation}
The ranges of coordinates $s, \theta,\phi $ are respectively given by $[-\infty,+\infty]\ , [0,\frac{\pi}{\mu}]\ , [0,+\pi]$. It is straightforward to check that these ranges cover the full \Poincare patch of AdS$_3$, {\it i.e.}~$\tau\in [-\infty, \infty]$, $x\in [-\infty, \infty]$ and $z \in [0, \infty]$. The metric in (\ref{m1}) describes an AdS$_3$ foliated by AdS$_2$ geometries at each $\phi={\rm const}$. It is instructive to note that, by comparing \eqref{m1} with eqn (3.1) in \cite{Spradlin:1999bn}, the constant $\phi$ slices correspond to global-$\text{AdS}_2$ geometry. We will revisit this in detail later.

    \vskip 0.5cm

\noindent   \textbf{\underline{For heating phase (d$<0$)}}:

When $d<0$, the set of equations in (\ref{em1})-(\ref{em3}) is solved by
    \begin{align}\label{a5}
        x & = \frac{\sqrt{d}}{2}\Bigl(\tan[\mu(s+i\theta)]+\tan[\mu(s-i\theta)]\Bigl)\ , \nonumber\\
        \tau & =\frac{\sqrt{d}}{2 i\sqrt{1+c_1^2}}\Bigl(\tan[\mu(s+i\theta)]-\tan[\mu(s-i\theta)]\Bigl) \ , \\
        z & =\frac{\sqrt{d} \ c_1}{2 i\sqrt{1+c_1^2}}\Bigl(\tan[\mu(s+i\theta)]-\tan[\mu(s-i\theta)]\Bigl)\ . \nonumber
    \end{align}
As before, substituting \eqref{a5} in AdS$_3$-\Poincare metric, $ds^2=\frac{dx^2+d\tau^2+dz^2}{z^2}$ we get:
    \begin{equation}\label{me2}
        ds^2=\frac{d\phi^2}{{\sin}^2[\phi]}+\frac{4 \beta^2 d\  (ds^2+d\theta^2)}{{\sin}^2[\phi]\ {\sinh}^2[2\sqrt{d}\beta\theta]}     \ .
    \end{equation}
    Again, the analytic continuation: $s\rightarrow is$ gives
    \begin{equation}\label{m2}
        ds^2=\frac{d\phi^2}{{\sin}^2[\phi]}+\frac{4 \beta^2 d\  (-ds^2+d\theta^2)}{{\sin}^2[\phi]\ {\sinh}^2[2\sqrt{d}\beta\theta]} \ .
    \end{equation}
    The ranges of variables $s, \theta,\phi $, for the metric \eqref{m2} in heating phase are given by $(-\infty,+\infty)\ , (0,\infty)\ , (0,\pi)$ respectively. The $\phi={\rm const}$ slices of (\ref{m2}) are now AdS$_2$ black holes which is explicitly visible by comparing (\ref{m2}) with equation (3.3) in \cite{Spradlin:1999bn}.

    \vskip 0.5cm
\noindent   \textbf{\underline{On the transition line (d$=0$)}}:

    By using exact similar analysis for $d=0$, the coordinates and  corresponding analytically continued metric can be written down as follows:
    \begin{align}\label{a6}
        x =- \frac{1}{2\beta}&\Bigl(\frac{1}{s+i\theta}+ \frac{1}{s-i\theta}\Bigl) \ , \quad
        \tau  =-\frac{1}{2 i\beta\sqrt{1+c_1^2}}\Bigl(\frac{1}{s+i\theta}- \frac{1}{s-i\theta}\Bigl)\ , \nonumber\\
        & z  =\frac{c_1}{2i\beta\sqrt{1+c_1^2}}\Bigl(\frac{1}{s+i\theta}- \frac{1}{s-i\theta}\Bigl) \ .
    \end{align}
    This yields:
     \begin{equation}\label{m3}
        ds^2=\frac{d\phi^2}{{\sin}^2[\phi]}+\frac{ -ds^2+d\theta^2}{{\sin}^2[\phi]\ {\theta}^2} \ .
    \end{equation}
In this case, the $\phi={\rm const}$ slices corresponds to the $\text{AdS}_2$-\Poincare patch\cite{Spradlin:1999bn}. We will now discuss how these patches determine the physics of the transition, especially by inserting explicit brane degrees of freedom inside the bulk geometry.

\numberwithin{equation}{section}
\section{Brane embeddings in AdS$_3$}
\label{bi1}

In this section we will demonstrate how a non-trivial conformal
boundary can detect the phase transition. Our explicit calculations
will be carried out in the Holographic description, since it
provides us with a natural and simple way to characterize various
boundary conditions on the conformal boundaries of the CFT. In the
Holographic dual, such boundaries correspond to defect branes which
are described by hypersurfaces in the geometry. We will show below
that these branes can detect the heating to non-heating phase
transition in both a probe limit as well as away from the probe
limit. Before proceeding further, let us recall that the relevant
metric data, in the Euclidean description, are given by
\begin{eqnarray}\label{me1}
&& ds^2=\frac{d\phi^2}{{\sin}^2[\phi]}+\frac{4 \mu^2 \  (ds^2+d\theta^2)}{{\sin}^2[\phi]\ {\sin}^2[2\mu \theta]} \ , \quad \mu = \beta \sqrt{|d|} \ , \quad d>0 \ , \\
&& s \in [-\infty, \infty] \ , \quad \theta \in [0, \frac{\pi}{\mu}] \ , \quad \phi \in [0, \pi] \ .
\end{eqnarray}
for the non-heating phase. Similarly, for the heating phase, we obtain:
\begin{eqnarray}\label{me2}
&& ds^2=\frac{d\phi^2}{{\sin}^2[\phi]}+\frac{4 \mu^2 \  (ds^2+d\theta^2)}{{\sin}^2[\phi]\ {\sinh}^2[2\mu \theta]} \ , \quad \mu = \beta \sqrt{|d|} \ , \quad d<0 \ , \\
&& s \in [-\infty, \infty] \ , \quad \theta \in [0, \infty] \ , \quad \phi \in [0, \pi] \ ,
\end{eqnarray}
Note that, in both (\ref{me1}) and (\ref{me2}), we can absorb the
factor of $\mu$ by redefining $s \to 2\mu s$ and $\theta \to 2 \mu
\theta$ and the resulting metric becomes independent of $\mu$.
However, the geometries retain the memory of ${\rm sgn}(d)$ since
(\ref{me2}) is obtained by sending $\mu \to - i \mu$ (equivalent to
sending $d \to -d$) in (\ref{me1}).\footnote{Recall that the phase
transition takes place as a function of ${\rm sgn}(d)$, which is the
sign of the Casimir of the $SU(1,1)$ evolution in the driven CFT.}
In the subsequent discussions, we will keep the factor of $\mu$
explicit.

\subsection{A Lorentzian Discussion}
\label{ld1}

\begin{figure}
    \centering
    \includegraphics[width=0.70\linewidth]{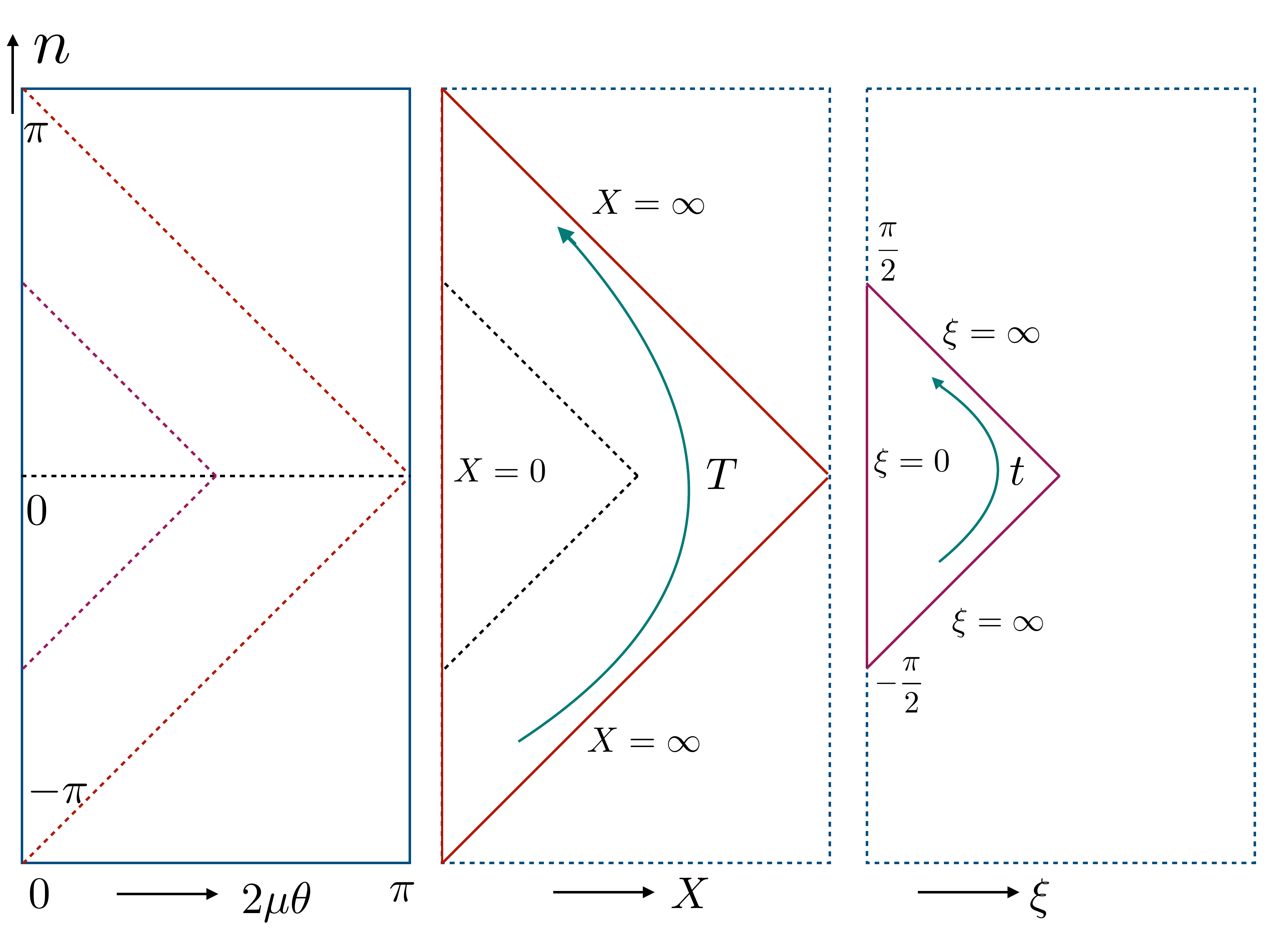}
    \caption{A pictorial representation of various patches of AdS$_2$ covered by various phases. The left-most is the non-heating phase that covers the global patch of AdS$_2$, the middle one covers the Poincar\'{e} patch of AdS$_2$ and the right most covers a Schwarzschild-like patch in AdS$_2$. These patches are explicitly related by the coordinate transformations in (\ref{nhtotr}) and (\ref{nhtoh}). }
    \label{fig:phasepatch}
\end{figure}

It is evident from (\ref{me1}) and (\ref{me2}) that the $\phi={\rm const}$ slices are special. This will prove crucial in the subsequent discussions and here we will discuss the Lorentzian picture in some detail, which will form the basic intuition in all subsequent observations. The Lorentzian patches are obtained by sending $s \to i s$ on the $\phi= \phi_0$ slices. The induced metric on the various phases are:
\begin{eqnarray}
&& ds^2 = \frac{4\mu^2}{\sin^2\phi_0} \left( \frac{- ds^2 + d\theta^2 }{\sin^2(2\mu\theta)}\right)  \ , \quad s\in [-\infty, \infty] \ , 2\mu \theta \in [0,\pi] \ ,\label{ads2nh} \\
&& ds^2 = \frac{4\mu^2}{\sin^2\phi_0} \left( \frac{- dt^2 + d \xi^2 }{\sinh^2(2\mu\xi)}\right) \ , \quad t \in [-\infty, \infty] \ , \xi \in [0,\infty]  \ , \label{ads2h} \\
&& ds^2 = \frac{4\mu^2}{\sin^2\phi_0} \left( \frac{- dT^2 + dX^2 }{X^2} \right) \ , \quad T \in [-\infty, \infty] \ , X \in [0,\infty] \ . \label{ads2tr}
\end{eqnarray}
Here (\ref{ads2nh}), (\ref{ads2h}) and (\ref{ads2tr}) correspond to non-heating, heating phases and at the transition point. These patches describe various parts of an AdS$_2$ geometry. Making explicit use of these metrics in \cite{Spradlin:1999bn}, they are also related to each other by simple co-ordinate transformations. Explicitly,
\begin{eqnarray}
&& T + X =  \tan \left(  \frac{s + 2\mu \theta}{2} \right) \ , \quad T-X = \tan \left(  \frac{s - 2\mu \theta}{2} \right) \ , \label{nhtotr} \\
&& \tan  \left( \frac{s - \frac{\pi}{2} + 2\mu\theta}{2} \right) = - e^{- 2\mu (t + \xi)} \ , \quad \tan  \left( \frac{s + \frac{\pi}{2} - 2\mu\theta}{2} \right) =  e^{ 2\mu (t - \xi)} \ , \label{nhtoh}
\end{eqnarray}
which relate the non-heating patch to the transition patch in
(\ref{nhtotr}) and the non-heating patch to the heating patch in
(\ref{nhtoh}). Note from (\ref{nhtotr}) that the line $s=2\mu\theta
-\pi$ and $s=\pi-2\mu\theta$ both map to $X=\infty$. On the other
hand, $\theta=0$ is mapped to $X=0$. Similarly, it is
straightforward to check that $s=2\mu\theta - \pi/2$ and $s=\pi/2 -
2\mu\theta$ map to $\xi=\infty$ line, while $\theta =0$ maps to
$\xi=0$ line. These patches are pictorially represented in
Fig.~\ref{fig:phasepatch}.
These $\phi = {\rm const}$ AdS$_2$ patches will be crucial in the subsequent sections.

\subsection{Probe Branes}
\label{pb1}

Let us first consider probing the geometries in (\ref{me1}) and (\ref{me2}) with a brane.\footnote{Note that probe branes can be used to detect phase transitions across a wide range of systems, {\it e.g.}~\cite{Albash:2006ew},\cite{Johnson:2008vna},\cite{Alam:2012fw}.} Consider a two-dimensional hypersurface with the following action:
\begin{eqnarray}
S_{\rm brane}=T\int d^2\sigma \sqrt{\gamma} = T \int d^2\sigma \cL \ ,
\end{eqnarray}
where, $\gamma_{ab}=g_{\mu\nu}\partial_a X^\mu\partial_b X^\nu$, is the induced metric and, $g_{\mu\nu}$ is the background metric and $T$ is the tension in the brane. Let us choose the world-volume coordinates to be: $\sigma^0 = s$, $\sigma^1=\theta$, and let $\phi(\theta)$ denote the corresponding embedding function. The corresponding induced metrics are:
\begin{eqnarray}
&& ds^2  =\frac{1}{\sin^2\phi(\theta)}\biggl(\frac{4\mu^2 ds^2}{\sinh^2(2\mu\theta)}+\left({\phi'}^2+\frac{4\mu^2}{\sinh^2(2\mu\theta)}\right)d\theta^2\biggl) \ , \quad {\rm heating} \ ,\\
&& ds^2  =\frac{1}{\sin^2\phi(\theta)}\biggl(\frac{4\mu^2 ds^2}{\sin^2(2\mu\theta)}+ \left ({\phi'}^2+\frac{4\mu^2}{\sin^2(2\mu\theta)}\right ) d\theta^2\biggl)  \ , \quad {\rm non-heating} \ .
\end{eqnarray}

With the above ansatz, the Lagrangian becomes a functional of the embedding function $\cL = \cL[\theta, \phi, \phi']$ and the brane profile can be obtained by solving the Euler-Lagrange equation:
\begin{eqnarray}
\frac{d}{d\theta} \left( \frac{\partial \cL}{\partial \phi'}\right) - \frac{\partial \cL}{\partial \phi} = 0 \ . \label{probebrane}
\end{eqnarray}
It is straightforward to observe that in both phases, the Euler-Lagrange equation admits a simple, analytical solution $\phi(\theta) = \pi/2$.\footnote{There is a family of solutions to the Euler-Lagrange equation, subject to appropriate boundary conditions, which can be obtained numerically.} For analytical control on the calculations, we will discuss only this solution. The corresponding on-shell actions of the probe branes, in both phases, can be computed by substituting this solution into the action. This yields:
\begin{eqnarray}
&& S^{\rm heating}_{\rm brane} =T \int ds d\theta \frac{4\mu^2}{\sinh^2(2\mu\theta)} = - \left. 2 \mu T \coth(2\mu\theta) \right|_{\theta_{\rm min}}^{\theta_{\rm max}} \int ds = \int ds \left( - 2 \mu T + \frac{T}{\epsilon_{\rm h}} \right)  \ , \\
&& S^{\rm non-heating}_{\rm brane} =T \int ds d\theta \frac{4\mu^2}{\sin^2(2\mu\theta)} = - \left. 2\mu T \cot(2\mu\theta)\right|_{\theta_{\rm min}}^{\theta_{\rm max}} \int ds = \int ds \left(\frac{2T}{\epsilon_{\rm nh}} \right) \ .
\end{eqnarray}
It is straightforward to observe that by choosing $\epsilon_{\rm h} = \epsilon_{\rm nh}/2$ the divergent pieces in the heating and the non-heating phases become equal. Here, to regulate the divergences, we have introduced two cut-offs $\epsilon_{\rm h} = \theta_{\rm min}$ in the heating phase, and $\epsilon_{\rm nh} = \theta_{\rm min} = \pi/(2\mu)  - \theta_{\rm max}$ in the non-heating phase.

The phase transition can be detected by considering the difference in their respective on-shell actions: $\Delta S = S^{\rm heating}_{\rm brane} - S^{\rm non-heating}_{\rm brane}$. This is formally divergent, unless we choose $\epsilon_{\rm h} = \epsilon_{\rm nh}/2$. This is certainly an allowed choice and it yields: $\Delta S \sim - 2 \mu T< 0$, $\forall \, \, T >0$. Alternatively, we can renormalize the corresponding on-shell actions by adding appropriate counter-terms to the respective branes. In the non-heating phase, there are two boundaries: $\theta \to 0$ and $\pi/ (2\mu) - \theta \to 0$, while the heating phase has only one boundary limit $\theta \to 0$. The corresponding on-shell action can be renormalized by introducing the following boundary terms:
\begin{eqnarray}
&& S_{\rm ct}^{\rm non-heating} = \left. \int ds \sqrt{h} \right|_{\theta = \frac{\pi}{\mu}-\epsilon_{\rm nh}} - \left. \int ds \sqrt{h} \right|_{\theta = \epsilon_{\rm nh}} \ , \\
&& S_{\rm ct}^{\rm heating} = - \left. \int ds \sqrt{h} \right|_{\theta = \epsilon_{\rm h}} \ ,
\end{eqnarray}
such that $S^{\rm heating}_{\rm brane}+S_{\rm ct}^{\rm heating}$ and $S^{\rm non-heating}_{\rm brane}+S_{\rm ct}^{\rm non-heating}$ are both finite. Here $h$ denote the induced metric on the boundary ({\it i.e.}~$\theta = {\rm const}$ slice) of the brane.

Several comments are in order. First, it is clear that for a fixed tension brane, the free energy is lowered as ${\rm sgn}(d)$ crosses zero from the positive side. We emphasize again that even though the factor of $(\mu T)$ can be absorbed in redefining $s$, the memory of ${\rm sgn} (d)$ remains in the final answer. Here ${\rm sgn}(d)$ corresponds to the sign of the Casimir that distinguishes between the non-heating and the heating phases.  This phase transition is a first order one, since it is straightforward to observe that {\it e.g.}~$\left( \partial S/ \partial \mu \right)$ have a discontinuous jump at the transition.\footnote{Also note that, the free energy does not have the detailed swallow-tail structure associated with a typical first order phase transition. This is perhaps due to the simplicity of the system. } Intuitively, one would prefer the positive tension branch, since it corresponds to a positive kinetic energy for the brane and satisfy standard positive energy conditions on the brane. The negative tension, on the other hand, corresponds to a negative kinetic energy and can lead to instabilities. Nonetheless, such objects appear naturally within the
context of string theory as {\it e.g.}~orientifold planes (see {\it e.g.}~\cite{Sen:1997kz}) and can play pivotal role in realizing interesting cosmological scenario.

\subsection{End-of-World (EOW) Branes}
\label{eow1}

We will now consider introducing fully back-reacting and dynamical End-of-World (EOW) branes in the corresponding heating and non-heating patches of AdS$_3$ geometry. Since the two patches are related by local co-ordinate transformations and not by large gauge transformations, the on-shell action of the three-dimensional Einstein-Hilbert term along with the Gibbons-Hawking boundary term cannot distinguish between the two phases. An EOW brane introduces a hypersurface dynamics which is determined by the extrinsic curvature and is therefore not a topological quantity in two-dimensions. Thus, it is expected that the phase transition will be explicitly visible once EOW-branes are inserted into the geometry. We will first consider a single EOW-brane and subsequently discuss two EOW-branes.

\subsubsection{Single EOW Brane}

The full bulk action now comprises of several pieces: The gravity part, the brane part and the intersection boundary part between the bulk geometry and the brane:
\begin{eqnarray}
&& S_{\rm full} = S_{\rm gravity} + S_{\rm brane} + S_{\rm corner} \ , \label{eowbranefull} \\
&& S_{\rm gravity} = - \frac{1}{2\kappa^2} \int_{\cM} d^3x \sqrt{g} \left( R - 2 \Lambda \right)  - \frac{1}{\kappa^2} \int_{\partial\cM} d^2y \sqrt{h} K  \ , \\
&& S_{\rm brane} = - \frac{1}{\kappa^2} \int_{\Sigma} d^2\sigma \sqrt{\gamma}  \left( K - T\right) \ , \\
&& S_{\rm corner} = - \frac{1}{\kappa^2} \int_{\cC} d\xi \sqrt{h_{\cC}} \left( \pi - \Theta_{\Sigma, \partial \cM}\right) \ , \quad \cC = \Sigma \cap \partial\cM \ .
\end{eqnarray}
Here $d^3x$, $d^2y$, $d^2\sigma$ and $d\xi$ denote the volume element of the full bulk geometry, the conformal boundary, the brane and the corner. Correspondingly, $g$, $h$, $\gamma$ and $h_{\cC}$ denote the metrics on them, $K$ denotes the corresponding extrinsic curvatures and $T$ is the brane tension. The angle $\Theta_{\Sigma, \partial \cM}$ denote the angle at which the brane intersects the conformal boundary. The variational problem on (\ref{eowbranefull}) is defined by varying the inverse metric within the region bounded by the branes and the boundary, keeping the branes and the corners fixed. This yields the following equations:
\begin{eqnarray}
&& R_{\mu\nu} - \frac{1}{2} \left( R - 2 \Lambda\right) g_{\mu\nu} = 0  \ , \label{einstein} \\
&& K_{ab} - \left( K - T \right)\gamma_{ab} = 0 \ . \label{brane}
\end{eqnarray}
The first equation ({\it i.e.}~Einstein equations) above determines the three-dimensional bulk geometry and the second equation determines the profile of the brane. For us, the Einstein equations are satisfied simply because we consider an AdS$_3$ geometry.

Before proceeding further, let us note that the non-heating patch is described by
\begin{eqnarray} \label{nonheat}
&& x = - \frac{\sqrt{|d|}}{2} v \ , \quad \tau = - \frac{\sqrt{|d|}}{2} u \cos \phi \ , \quad z= - \frac{\sqrt{|d|}}{2} u \sin\phi \ , \\
&& u = \frac{1}{i} \left[ \coth\left(\mu (s + i \theta) \right) - \coth\left(\mu (s - i \theta) \right)\right] \ , \quad v = \left[ \coth\left(\mu (s + i \theta) \right) + \coth\left(\mu (s - i \theta) \right)\right] \ . \nonumber\\
\end{eqnarray}
Similarly, the heating phase is described by the following patch:
\begin{eqnarray}\label{heat}
&& x = - \frac{\sqrt{|d|}}{2} v \ , \quad \tau =  \frac{\sqrt{|d|}}{2} u \cos \phi \ , \quad z=  \frac{\sqrt{|d|}}{2} u \sin\phi \ , \\
&& u = \frac{1}{i} \left[ \tan\left(\mu (s + i \theta) \right) - \tan\left(\mu (s - i \theta) \right)\right] \ , \quad v = \left[ \tan \left(\mu (s + i \theta) \right) + \tan \left(\mu (s - i \theta) \right)\right] \ . \nonumber\\
\end{eqnarray}
Both the patches are essentially described by the following equations:
\begin{eqnarray}
&& \tau^2 + z^2 = \frac{d}{4} u^2 \ , \label{circular} \\
&& \frac{z}{\tau} = \tan\phi \ . \label{line}
\end{eqnarray}
From (\ref{circular}), we observe that each $u={\rm const}$ describes a leaf of a circular foliation of the Poincar\'{e} patch and (\ref{line}) implies that each $\phi = {\rm const}$ describes a leaf of a planar foliation of the same. It is expected that each leaf of both the planar and the circular foliation is described by a Karch-Randall brane of a given tension. We will explicitly show this and for simplicity, we will focus on the planar foliations.
\begin{figure}
    \centering
    \includegraphics[width=0.70\linewidth]{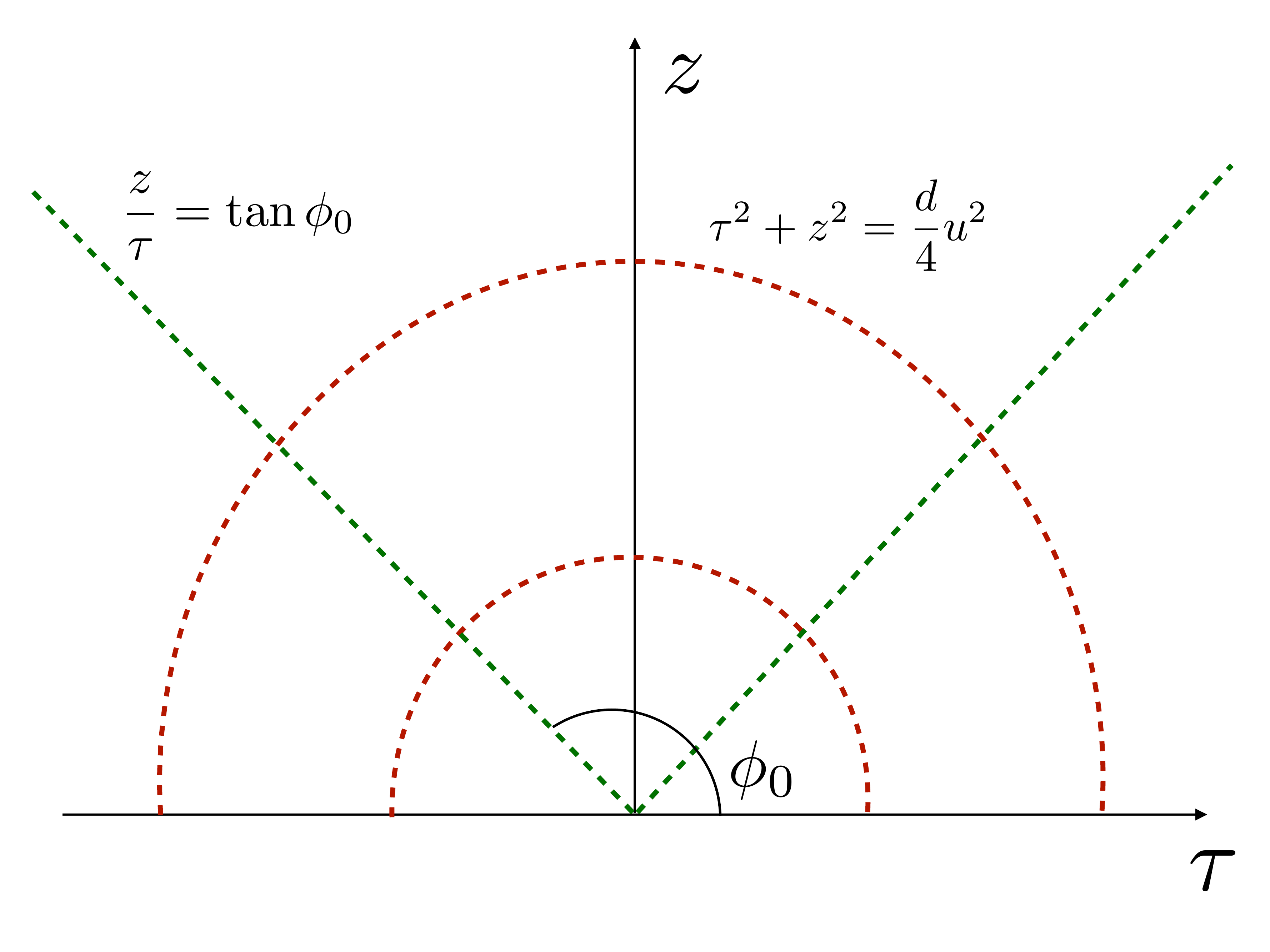}
    \caption{A pictorial representation of the two types of foliations described by equations (\ref{circular}) and (\ref{line}). For the linear leaves, positive tension Karch-Randall branes are located at constant angle $\phi_0 \in [\frac{\pi}{2}, \pi]$ and for negative tension, they correspond to $\phi_0\in [0,\frac{\pi}{2}]$. Here, the EOW-brane intersects the conformal boundary at $\tau=0$. }
    \label{fig:foliations}
\end{figure}

Let us choose $\sigma^0 = s$ and $\sigma^1 = \theta$ as the worldvolume coordinates, and $\phi= \phi(\theta)$ as the embedding function. This implies: $d\phi - \phi' d\theta =0$ and therefore
the unit outward normal to the brane is given by
\begin{eqnarray}
&& n_\alpha^{\rm nh} = \frac{2 \mu}{\sin\phi}\frac{1}{\sqrt{4 \mu^2 + \phi'^2 \sin^2(2 \mu\theta)}}\left( 0, - \phi', 1\right) \ , \\
&& n_\alpha^{\rm h} = \frac{2 \mu}{\sin\phi}\frac{1}{\sqrt{4 \mu^2 + \phi'^2 \sinh^2(2 \mu\theta)}}\left( 0, - \phi', 1\right) \ ,
\end{eqnarray}
where $n_\alpha^{\rm nh}$ and $n_\alpha^{\rm h}$ correspond to non-heating and heating phases, respectively. The extrinsic curvature is calculated by using $K_{ab} = \nabla_\alpha n_\nu e_a^\alpha e_b^\nu$, where
$e_a^\alpha= \left( \partial x^\alpha/\partial \sigma^a \right)$. The simplest component of the brane equation in (\ref{brane}) is given by the $nn$-component which can be readily solved to obtain the profile: $\phi(\theta) = \phi_0$, where $T = - \cos \phi_0$, in both phases. It is now straightforward to check that this solves the full equations in (\ref{brane}).

Let us now evaluate the on-shell actions in the corresponding phases. First, in the non-heating phase, we obtain:
\begin{eqnarray}
&& S_{\rm gravity} = - \frac{1}{\kappa^2} \frac{1}{\epsilon^2} \int ds \int_0^{\pi/2\mu} \frac{d\theta}{\sin^2(2\mu \theta)} \ , \quad S_{\rm brane} = -\frac{1}{\kappa^2} \frac{2}{\epsilon} \frac{T}{1-T^2} \int ds \ , \\
&& S_{\rm corner} = - \frac{1}{\kappa^2}   \phi_0 \frac{1}{\epsilon}\int ds \ ,
\end{eqnarray}
We add the following counter-term:
\begin{eqnarray}
&& S_{\rm ct} = \left. \frac{1}{\kappa^2} \left( \frac{2T}{1-T^2} + (\pi - \phi_0) \right)  \int d\xi \sqrt{h} \right|_{\theta=\epsilon} +  \left. A_{\rm nh} \frac{1}{\kappa^2} \int \sqrt{h}  \right|_{\phi=\pi- \epsilon} \ ,\\
&& A_{\rm nh} = \int dn \int_0^{\pi/2\mu} \frac{d\theta}{\sin^2(2\mu \theta)} \ .
\end{eqnarray}
Here $h$ denotes the induced metrics at the corresponding hyper-surfaces. Note that the $z = \epsilon$ hypersurface corresponds to $\phi = \pi-\epsilon$ hypersurface. Also note that, the coefficient $A_{\rm nh}$ is formally a divergent quantity, which itself needs a regularization. Nonetheless, the upshot is that there is no finite contribution from the counter-terms and therefore the sum of $S_{\rm full} + S_{\rm ct} =0$, in the non-heating phase.

A similar computation in the heating phase require identical counter-terms as above, except $A_{\rm nh} \to A_{\rm h}$, where
\begin{eqnarray}
A_{\rm h} = \int ds \int_0^{\infty} \frac{d\theta}{\sinh^2(2\mu \theta)} \ .
\end{eqnarray}
Note that, while $A_{\rm h}$ is still formally a divergent quantity and needs regularization, the $\theta$-integral produces a finite contribution as $\theta \to \infty$. This very feature becomes crucial on the brane. In the non-heating phase, the brane on-shell action consists only of divergent contributions while in  the heating phase the $\theta$-integral contains a finite piece, as we just noticed. Upon introducing the counter-terms this finite contribution survives and we obtain:
\begin{eqnarray} \label{oneeow}
S_{\rm full} + S_{\rm ct}  = \frac{1}{\kappa^2} \frac{2\mu T}{1-T^2} \int ds \ .
\end{eqnarray}
A few comments are in order. Note that, in the tension-less limit $T\to 0$ and therefore $\phi_0 \to \pi/2$, which recovers the probe limit answer of equation (\ref{probebrane}). In the strict $T=0$ limit, (\ref{oneeow}) vanishes, which is also consistent with the probe limit calculation. The free energy in the small tension limit, however, does not reduce to the probe limit answer since the extrinsic curvature still contributes to the full action. This is manifest in (\ref{oneeow}), in which $\Delta S >0$, whereas the probe calculation yields $\Delta S <0$. Nevertheless, in both calculations, the phase transition is detectable. A final comment is on the discontinuity of the first derivative of the free energy across the transition. This is obtained by computing $(\partial S)/(\partial d) \sim d^{-1/2} \to \infty$\footnote{Recall that $\mu = \beta \sqrt{|d|}$.} as $d\to 0$. Hence the phase transition is accompanied by a generally divergent discontinuity, except in a fine-tuned limit $T\to 0$, in which it can become a finite quantity. Alternatively, we can consider a derivative with respect to $\mu$, which will remain finite and the corresponding phase transition will be associated with a finite discontinuity of the derivative. Finally note that, as $T\to 1$, the free energy and all other associated physical quantities diverge. This is a singular limit, in which the EOW brane coincides with the conformal boundary of AdS and cuts-off the entire geometry.

Let us now briefly discuss the dual CFT picture. The insertion of an
EOW-brane in the bulk amounts to introducing a boundary in the dual
CFT, following the proposals in \cite{Takayanagi:2011zk,
Fujita:2011fp}. These boundaries preserve conformal symmetries and
the corresponding boundary states are obtained by solving $\left(L_p
- \bar{L}_{-p}\right) |B\rangle =0$, where $\{L_p, \bar{L}_q\}$ are
the holomorphic and anti-holomorphic copies of Virasoro generators.
A general boundary state $|B\rangle$ can be constructed from a
linear combination of the so-called Ishibashi
states\cite{Ishibashi:1988kg}. As Fig.\ \ref{fig:foliations}
demonstrates, the CFT is defined on $x \in [-\infty, \infty]$ and
$\tau \in [0, \infty]$. The corresponding boundary state can be
labelled by an index $| B_\alpha \rangle$, which is encoded in the
tension of the brane. Subsequently, for a CFT defined on a cylinder,
the Euclidean on-shell action is related to the disc partition
function for the BCFT, given by $\langle 0 | B_\alpha \rangle \equiv
g_\alpha$. Note, however, that we started with a \Poincare AdS$_3$
geometry and therefore the Holographic on-shell action is not simply
related to the disc partition function. Instead, it computes the CFT
partition function defined on the half-plane in $\tau$. Finally, a
note of caution: Recall that the action of the bulk Hamiltonian as
well as the equations for the tangent curves are obtained starting
from a \Poincare AdS bulk. One can also begin with a bulk dual of a
BCFT and subsequently analyze the bulk Hamiltonian as well as the
tangent curves accordingly. This description contains an EOW-brane
to begin with and it will be interesting to analyze this case
further. We do not, however, expect any qualitative difference in
the physical picture.

\subsubsection{Two EOW Branes}

Let us now consider two such EOW-branes. The corresponding action is given by
\begin{eqnarray}
&& S_{\rm full} = S_{\rm gravity} + S_{\rm brane} + S_{\rm corner} \ , \\
&& S_{\rm gravity} = - \frac{1}{2\kappa^2} \int_{\cM} d^3x \sqrt{g} \left( R - 2 \Lambda \right)  - \frac{1}{\kappa^2} \int_{\partial\cM} d^2y \sqrt{h} K  \ , \\
&& S_{\rm brane} = \sum_{i=1,2} - \frac{1}{\kappa^2} \int_{\Sigma_i} d^2\sigma \sqrt{\gamma}  \left( K - T_i \right) \ , \\
&& S_{\rm corner} = - \frac{1}{\kappa^2} \int_{\cC} d\xi \sqrt{h_{\cC}}  \Theta_{\Sigma, \partial \cM} \ , \quad \cC = \left( \Sigma_1 \cap \partial\cM\right) \cup \left( \Sigma_2 \cap \partial\cM\right) \cup \left(\Sigma_1\cap \Sigma_2\right) \ .
\end{eqnarray}
Here $\Sigma_{i}$, $i=1,2$ denote the two EOW branes and $T_i$ are the corresponding tensions. The equations of motion are still given by (\ref{einstein}) and (\ref{brane}) and the corresponding brane solutions are $T_1  = - \cos\phi_1$ and $T_2 = - \cos\phi_2$. Now these two EOW-branes may intersect in the bulk, in which case there is a non-trivial finite contribution to the free energy coming from the intersection point of the two branes. However, if the two branes remain non-intersecting, then there is no such contribution and the result remains the same as above.

\begin{figure}
    \centering
    \includegraphics[width=0.70\linewidth]{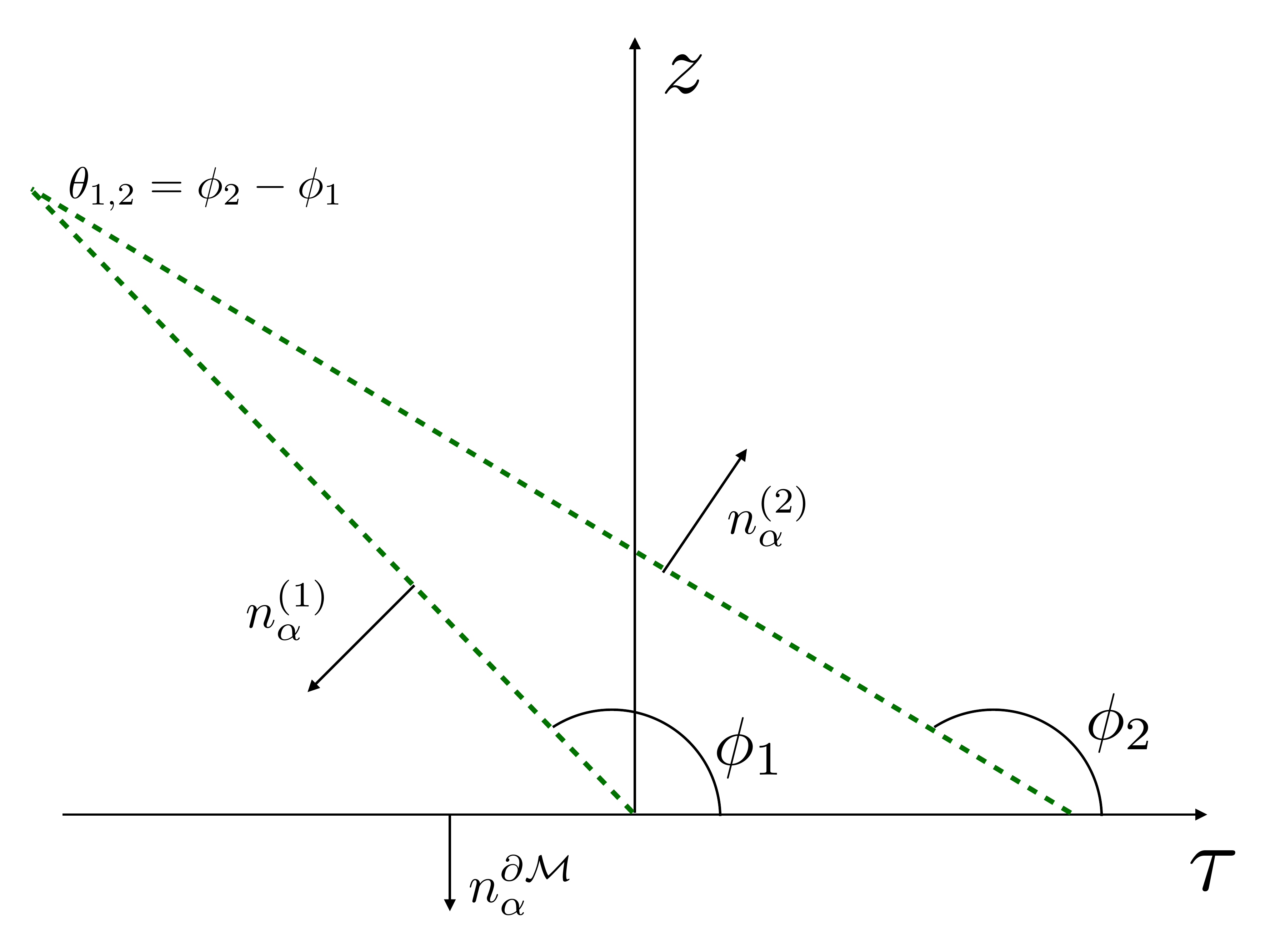}
    \caption{A pictorial representation of two intersecting EOW-branes. The outward normal to $\Sigma_{1,2}$ are denoted by $n_\alpha^{1,2}$ and the outward normal to the conformal boundary of AdS is denoted by $n_\alpha^{\partial \cM}$. The two branes intersect at an angle $\theta_{1,2}= \phi_2 - \phi_1$ in the bulk. The gravitational theory is defined within the triangle, including its sides and corners. }
    \label{fig:twoeow}
\end{figure}

The intersecting case is shown in Fig.~\ref{fig:twoeow}.
The analyses proceeds as above with one important addition. Since the outward normals at $\Sigma_1$ and $\Sigma_2$ satisfy: $g^{\alpha \beta} n_\alpha^{(1)} n_\beta^{(2)} <0 $, we choose:
\begin{eqnarray}
n_\alpha^{(1)} = \frac{1}{\sin\phi_1} \left( 0, 0, 1\right) \ ,  \quad n_\alpha^{(2)} = - \frac{1}{\sin\phi_2} \left( 0, 0, 1\right)  \ ,
\end{eqnarray}
in both heating and non-heating phases. Here, we have already used
the solution for the EOW-brane $\phi_{1,2}'=0$. From
Fig.~\ref{fig:twoeow}, the branes are intersecting if $\theta_{1,2}
= \phi_2 - \phi_1 > 0$;\footnote{This, in turn, implies that $\tau_1
< \tau_2$.} and there is no intersection if $\theta_{1,2} \le 0$. To
proceed further, it convenient to describe these branes in the
Poincar\'{e} patch:
\begin{eqnarray}
\frac{z}{\tau - \tau_1} = \tan\phi_1 \ , \quad \frac{z}{\tau-\tau_2} = \tan\phi_2 \ ,
\end{eqnarray}
where $\tau_{1,2}$ are the points at which the EOW-branes $\Sigma_{1,2}$ intersect the conformal boundary of AdS. Their mutual intersection point is given by
\begin{eqnarray}
\tau_* = \frac{\tau_2 \tan\phi_2 - \tau_1 \tan\phi_1}{\tan\phi_1 - \tan\phi_2} \ , \quad z_* = \tan\phi_1 \tan \phi_2 \frac{\tau_1-\tau_2}{\tan\phi_1 - \tan\phi_2} \ ,
\end{eqnarray}

It is easy to evaluate the pure-gravity part of the on-shell action in both phases. This yields:
\begin{eqnarray}
S_{\rm gravity} = \frac{1}{\kappa^2} (\tau_1 - \tau_2) \frac{1}{z_*^2}\int dx + \frac{C_1}{\epsilon^2} \ ,
\end{eqnarray}
which consists of either a divergent piece or a finite term that is
universal in both phases. Thus, this will not be relevant in the
free energy differences. Likewise, the intersection terms $S_{\rm
corner}$ are universal in both phases, except for the contribution
coming from the mutual intersection point of the two branes. To
proceed further, we now need to fix the ranges of coordinates
corresponding to the region enclosed by the branes and the conformal
boundary of AdS (see {\it e.g}~Fig.~\ref{fig:twoeow}).

The domain of interest is defined by the Poincar\'{e} coordinates $x
\in [-\infty, \infty]$, $\tau \in (\tau_*, \tau_2)$ and $z \in (0,
z_*)$. In the non-heating phase, recall that:
\begin{eqnarray}
\tau = - \frac{\sqrt{|d|}}{2} u \cos \phi_{1,2} \ , \quad z= - \frac{\sqrt{|d|}}{2} u \sin\phi_{1,2} \ , \quad u = \frac{1}{i} \left[ \coth\left(\mu (s + i \theta) \right) - \coth\left(\mu (s - i \theta) \right)\right] \ .
\end{eqnarray}
It is clear that $z_*=z_*(s, \theta)$, and therefore the
corresponding ranges of the coordinates $\{s, \theta\}$ are mutually
dependent. For example, setting $s=0$,\footnote{Recall that this
choice does not affect the stroboscopic time $n$ to be a suitably
large integer. This is simply because there is always a shift
freedom between these two coordinates.} the corresponding coordinate
ranges are given by
\begin{eqnarray}
 \theta \in \left [\theta_*,  \frac{\pi}{2\mu} \right] \ , \quad \theta_* = \frac{1}{\mu} \cot^{-1} \left( \frac{z_*} {\sqrt{|d|} \sin\phi_{1,2}}\right)  \ . \label{non_heatrange}
\end{eqnarray}
Similarly, in the heating phase, one obtains the following range:
\begin{eqnarray}
 \theta \in \left [\theta_*, 0 \right]  \ , \quad \theta_* = \frac{1}{\mu} \tanh^{-1}\left( \frac{z_*}{\sqrt{|d|} \sin\phi_{1,2}}  \right)   \ , \label{heatrange}
\end{eqnarray}
where we implicitly assume that the intersection point $z_*$ remains within the corresponding patch. Note that, the ranges in (\ref{non_heatrange}) and (\ref{heatrange}) both depend on $z_*$, given a brane angle $\phi_{1,2}$. In general, therefore, $\theta_*= \theta_*(s)$. Thus, the region bounded by the EOW-branes becomes explicitly dynamical. Correspondingly, the on-shell action also depends explicitly on $s$. For simplicity, we will be working with a {\it free energy density} defined at the $s=0$ slice, using the above ranges.

In the non-heating phase, the finite contribution from the brane and the corner part evaluates to:
\begin{eqnarray}\label{freenonheat}
&& S_{\rm brane}  = \frac{2\mu}{\kappa^2}  \left[ \frac{\cos\phi_1}{\sin^2\phi_1} \cot \left( 2 \cot^{-1} \left( \frac{z_*} {\sqrt{|d|} \sin\phi_{1}}\right) \right) + \frac{\cos\phi_2}{\sin^2\phi_2} \cot \left( 2 \cot^{-1} \left( \frac{z_*} {\sqrt{|d|} \sin\phi_{2}}\right) \right) \right] \nonumber\\
&& S_{\rm corner} = - \frac{2\mu}{\kappa^2}  \left( \phi_2 - \phi_1 \right) \csc \left( 2 \cot^{-1} \left( \frac{z_*} {\sqrt{|d|} \sin\phi_{1}}\right) \right) \ .
\end{eqnarray}
In the heating phase, the corresponding finite contributions are:
\begin{eqnarray}\label{freeheat}
&& S_{\rm brane}  =  \frac{2\mu}{\kappa^2}  \left[ \frac{\cos\phi_1}{\sin^2\phi_1} \coth \left( 2 \tanh^{-1} \left( \frac{z_*} {\sqrt{|d|} \sin\phi_{1}}\right) \right)  \right. \nonumber\\
&& \left.+ \frac{\cos\phi_2}{\sin^2\phi_2}  \coth \left( 2 \tanh^{-1} \left( \frac{z_*} {\sqrt{|d|} \sin\phi_{2}}\right) \right) \right] \nonumber\\
&& S_{\rm corner} = - \frac{2\mu}{\kappa^2}  \left( \phi_2 - \phi_1 \right) \left(\sinh \left( 2 \tanh^{-1} \left( \frac{z_*} {\sqrt{|d|} \sin\phi_{1}}\right) \right) \right)^{-1} \ .
\end{eqnarray}
The free energy difference is now given by taking the difference between (\ref{freeheat}) and (\ref{freenonheat}).

It is straightforward to check that in the special case when $T_2=0$ and $T_1=T$, we get back the same answer as in (\ref{oneeow}). In the special case when $T_1=T_2=T$, we also get back the same qualitative physics, since the free energies are enhanced by a factor of two, keeping the sign and the behaviour of the difference the same.
\begin{figure}
    \centering
    \includegraphics[width=0.70\linewidth]{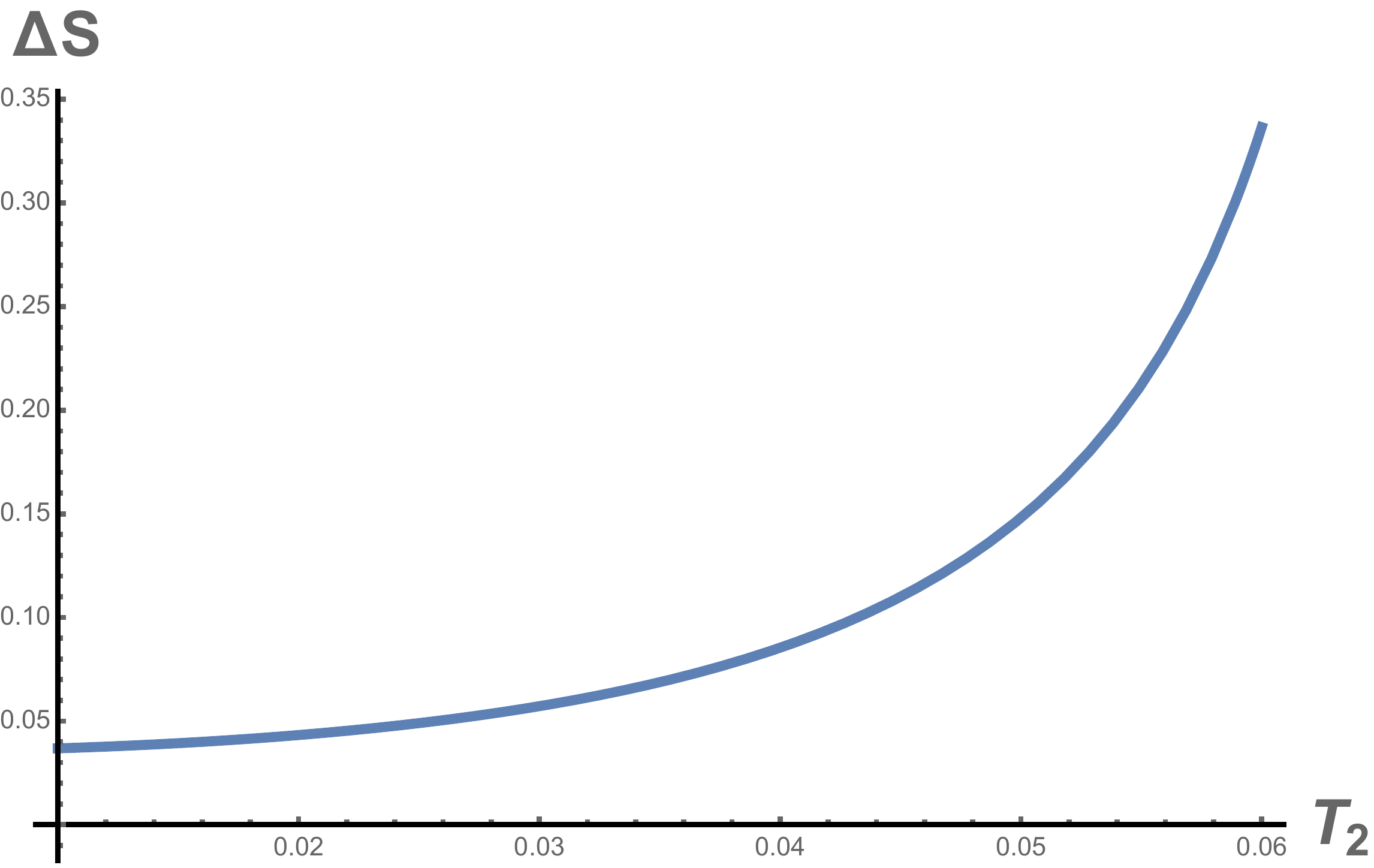}
    \caption{A representative behaviour of the free energy difference $\Delta S = S_{\rm heating} - S_{\rm non-heating}$, for fixed values of $T_1=0.001$, $|d|=2$, $\tau_1-\tau_2= -0.1$, within a specified range of values of $T_2$ which are shown in the figure. This plot shows a monotonically increasing and a positive $\Delta S$ in this range of $T_2$, with $\Delta S \to \infty$ as $T_2 \to T_2^c\approx 0.0715$. At $T_2=T_2^c$, there is an infinite jump. Thus, the non-heating phase, in this branch, has a lower free energy.}
    \label{fig:2phase1}
\end{figure}
\begin{figure}
    \centering
    \includegraphics[width=0.70\linewidth]{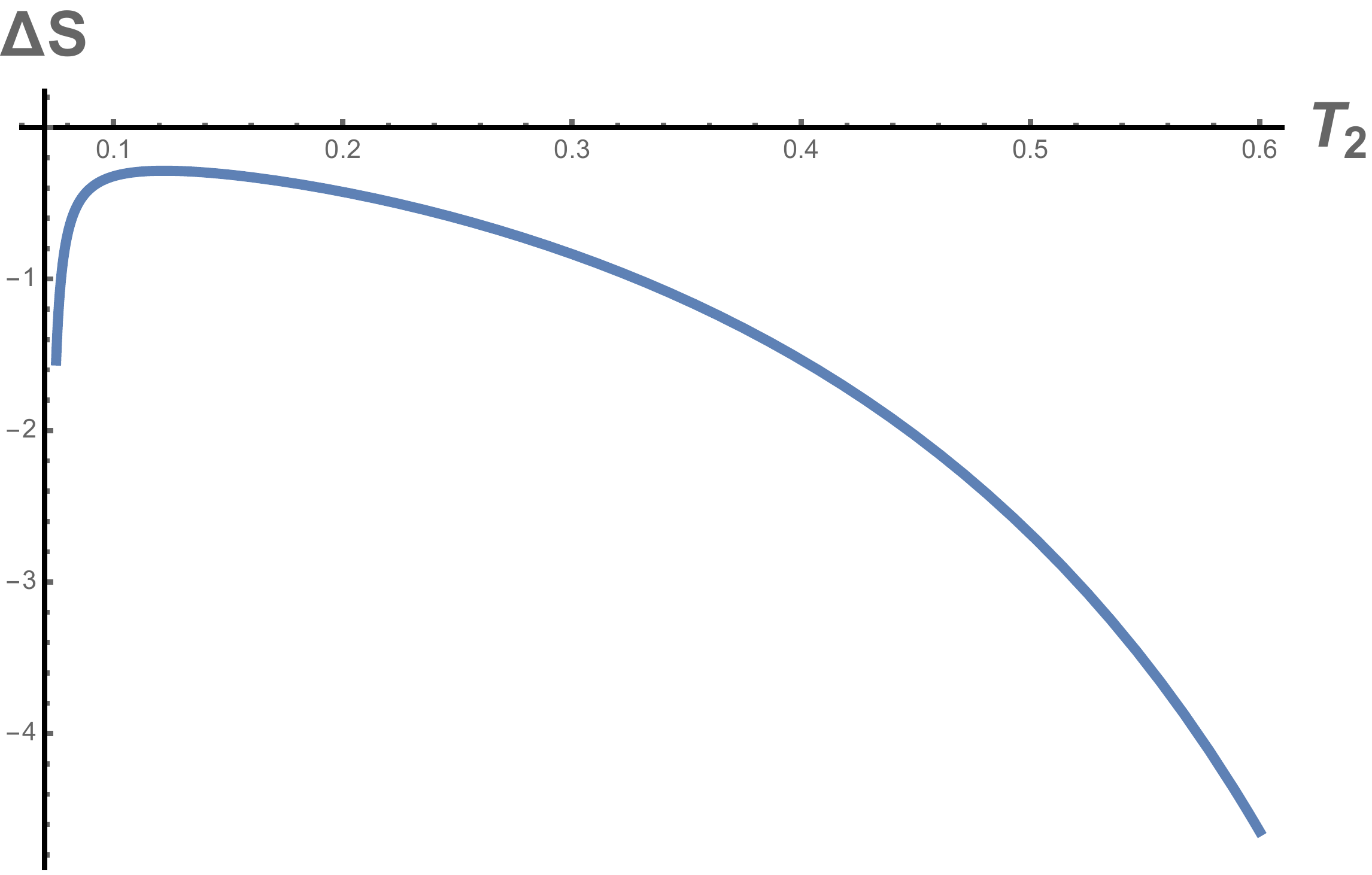}
    \caption{A representative behaviour of the free energy difference $\Delta S = S_{\rm heating} - S_{\rm non-heating}$, for fixed values of $T_1=0.001$, $|d|=2$, $\tau_1-\tau_2= -0.1$, within a specified range of values of $T_2$ which are shown in the figure. This plot shows a negative $\Delta S$ in this range of $T_2>T_2^c\approx 0.0715$. Thus, the heating phase has a lower free energy in this branch.}
    \label{fig:2phase2}
\end{figure}
The general behaviour is richer. A representative feature is shown
in Figs.~\ref{fig:2phase1} and \ref{fig:2phase2}. For a given $T_1$,
the free energy difference has two distinct signatures in two
regimes of $T_2$. These two regimes are demarcated by the point at
$z_* = \sqrt{|d|} \sin\phi_1$, which yields:
\begin{eqnarray}
\phi_2 = \arctan \left( \frac{\sqrt{|d|} \sin\phi_1} {\tau_1 - \tau_2 + \sqrt{|d|} \cos\phi_1}\right)  \ .
\end{eqnarray}
At this location $\Delta S \to \pm \infty$, as $\phi_2$ approaches the above value from above or from below.

\begin{figure}
    \centering
    \includegraphics[width=0.70\linewidth]{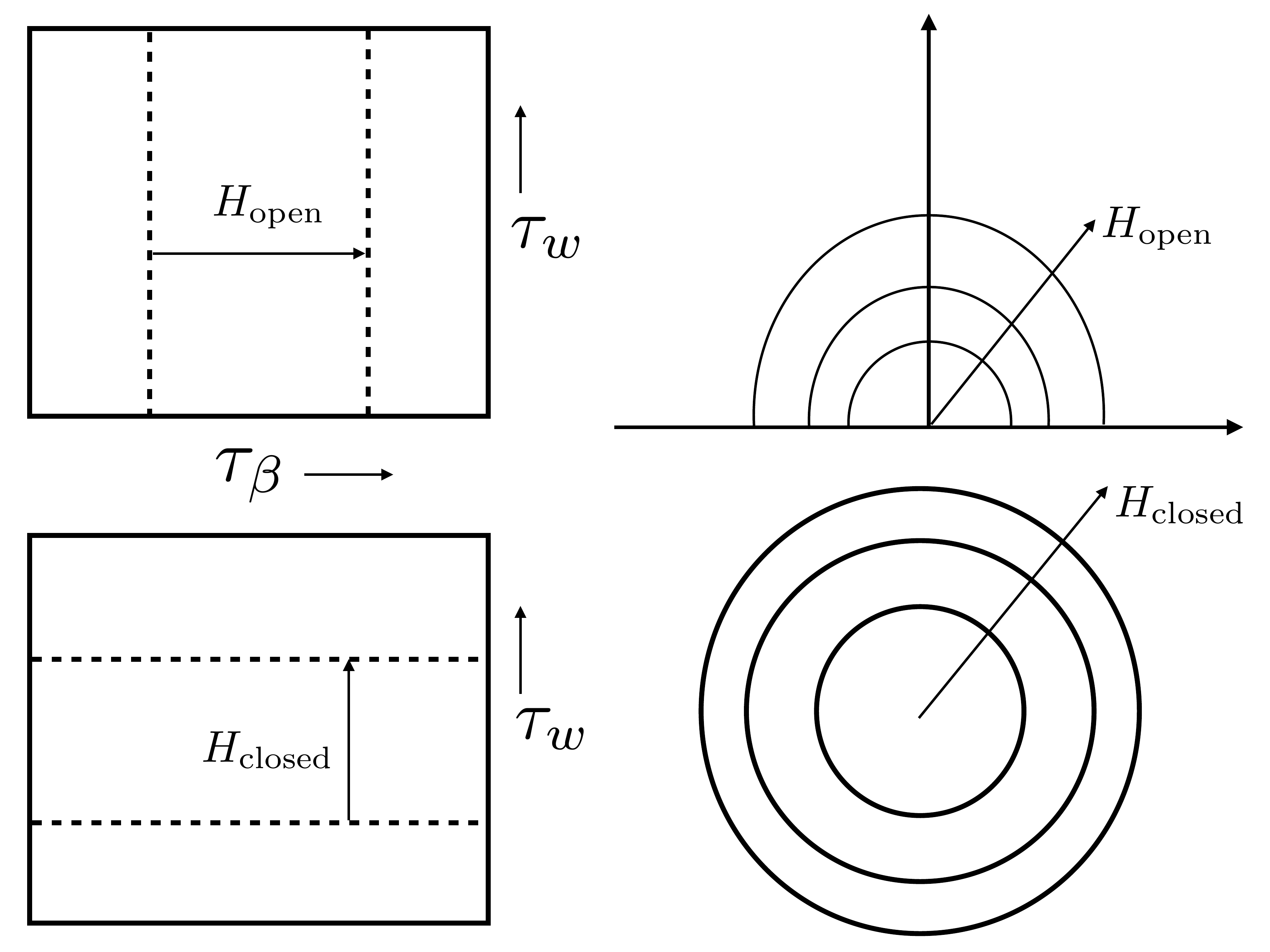}
    \caption{A pictorial representation of the BCFT partition function, which is defined on a rectangular region of horizontal length $\tau_\beta$ and a vertical length of $\tau_w$. Let us assume that $\tau_\beta$ is periodic. The picture above corresponds to the open string channel, in which an open string has end points at the two horizontal lines, separated by $\tau_w$. The corresponding CFT can be defined on the upper half plane, which is shown on the right. The picture below corresponds to the closed string channel, in which a closed string of circumference $\tau_\beta$ propagates from an initial state to a final state. The corresponding CFT is defined on the entire plane, with different states inserted on circles of different radii.}
    \label{fig:openclosed}
\end{figure}

Let us now discuss the dual CFT perspective. The presence of two
boundaries in the CFT has two different interpretations for the
corresponding BCFT. In the so-called open string channel, one
considers an open string with two end points at the two boundaries.
Alternatively, one can adopt a closed-string channel description, in
which case a closed string state evolves from an initial state to a
final state. Consider the Euclidean path integral, denoted by
$Z_{ab}$, of a CFT on a cylinder with circumference $\tau_\beta$ and
vertical width $\tau_w$, with boundary conditions $a$ and $b$ at the
two ends. See Fig.~\ref{fig:openclosed} for a pictorial
representation. As before, we add a note of caution: One can
alternatively begin with a bulk geometry with the EOW-branes already
inserted and explore the corresponding patches by analyzing the bulk
Hamiltonian and the corresponding tangent curves. We expect the key
qualitative aspect to remain unchanged, however, it is an
interesting scenario to explore in detail.

In the open-string channel, $Z_{ab}$ can be thought of as a thermal
partition function for a system defined within an interval of width
$\tau_w$ with boundary conditions $a$ and $b$ at the two end points.
Thus, $Z_{ab} = {\rm Tr} (e^{-\tau_\beta H_{\rm open}})$. In the
closed string channel, this becomes a transition amplitude between
two boundary states, $|a\rangle$ and $|b\rangle$, in a system which
is defined on a circle of circumference $\tau_\beta$. Thus, $Z_{ab}
= \langle a | e^{-\tau_w H_{\rm closed}}| b\rangle$. Note that this
geometry is characterized by the dimensionless ratio
$\tau_w/\tau_\beta$, up to its conformal class. It can be shown that
in the limit $\tau_w/\tau_\beta \to \infty$, the Euclidean path
integral is given by: $Z_{ab} = g_a g_b e^{\frac{\pi c
\tau_w}{6\tau_\beta}}$, where $c$ is the central charge of the CFT
and $g_{a,b} = \langle a, b | 0\rangle$. These $g_{a,b}$ are ground
state degeneracies. In the limit, $\tau_w/\tau_\beta \to \infty$,
the Holographic on-shell action is precisely related to these ground
state degeneracies. Note, however, that our bulk dual is based on
the \Poincare AdS$_3$ geometry and therefore the CFT is defined on a
decompactified circle: $\tau_\beta \to \infty$. In this limit, the
bulk on-shell action still computes the CFT partition function
$Z_{ab}$, and this receives contribution from ground state as well
as excited states. To precisely connect with boundary entropy of the
BCFT, one should begin with a global AdS$_3$ geometry and
subsequently carry out the analyses above. It is an interesting
aspect, which we leave for  a future work.

A final note is about the types of boundary conditions. With two
EOW-branes, {\it i.e.}~with two boundaries one can define a boundary
condition changing operator. These operators are formally defined as
the primary operators with the smallest dimension, in the spectrum
of open string channel with two non-identical boundary conditions at
the two end points. This is non-trivial when $a\not = b$, which
corresponds two EOW-branes with two different tensions $T_1\not =
T_2$.  As we have demonstrated above, this has a rich structure
associated with the phase transition.


\numberwithin{equation}{section}
\section{Boundary correlation functions from the bulk geometry}
\label{corr1}

In this section, we compute two-point and four-point correlation
functions in the bulk in all three phases. We then compare them with
the known results in a large $c$ CFT \cite{Das:2021gts,
Das:2022jrr1}. This will provide a self-consistency check on our
geometric description.

\subsection{Two-point correlation functions}

We will compute the two-point function using the geodesic approximation
\cite{Roberts:2014ifa1}\cite{Shenker:2013pqa1}, wherein the two
point function is approximated by the exponential of the geodesic
distance between the two boundary operators. To this end, we first
express the geodesic in terms of the embedding coordinates ($T_1$,
$T_2$, $X_1$, $X_2$) of $AdS_3$.

    %


 The geodesic distance ($D$) between two points whose embedding coordinates are ($T_1$, $T_2$, $X_1$, $X_2$) and ($T'_1$, $T'_2$, $X'_1$, $X'_2$), is given by\cite{Shenker:2013pqa1} \footnote{We have set the AdS radius $l=1$.}
    \begin{equation}\label{gl1}
        \cosh D=T_1 T'_1+T_2 T'_2-X_1 X'_1-X_2 X'_2 \ .
    \end{equation}

We then express the answer in terms of the bulk coordinates by using the explicit map between the bulk coordinates in which the bulk metric is written for each of the phases and the embedding coordinates. We do this for each phase separately and after regulating the divergence in the geodesic distance, we find an exact match with the boundary two-point function.


\subsubsection{2-point correlation function in the heating phase}\label{A}

    We start with the metric \eqref{me2} for the heating phase $(d<0)$ and consider the following coordinate changes:
    \begin{equation}\label{cb1}
        r=\coth(2\mu \theta) \ \text{and  } \ t=2\mu s \ ,
    \end{equation}
    to rewrite \eqref{me2} as:
    \begin{equation}\label{ma1}
        ds^2=\frac{d\phi^2}{\sin^2\phi}+\frac{1}{\sin^2\phi}\biggl(\frac{dr^2}{r^2-1}-(r^2-1)dt^2\biggl) \ .
    \end{equation}
The embedding coordinates corresponding to \eqref{ma1} are given by:
 \begin{align}\label{e1}
        T_1 & = \sqrt{r^2-1}\sinh t\csc\phi , \nonumber\\
        T_2 & = r \csc\phi , \nonumber\\
        X_1 & = \cot\phi  \ , \nonumber \\
         X_2 & = \sqrt{r^2-1}\cosh t\csc\phi .
\end{align}
We now compute the correlators of two boundary operators $V$\footnote{Here, the operators in consideration are heavy operators of mass $m$.} which are located at $(t_1,r_1,\phi=0)$ and $(t_2,r_2,\phi=0)$.

In this set-up, the geodesic length in $\eqref{gl1}$ turns out to be:
\begin{equation}\label{gl2}
    \cosh D=\lambda^2(r_1 r_2-1-\sqrt{(r_1^2-1)(r_2^2-1)}\cosh(t_1-t_2)) \ .
\end{equation}
Here $\lambda \equiv \csc\phi$. For the boundary points, this is actually divergent since the boundary points are at $\phi=0$. We therefore regulate it by taking $\lambda$ large but not infinite and then removing the divergent term to obtain the regularized geodesic distance. In this limit, the expression simplifies to:
\begin{equation}\label{gl2}
    D\sim\log\biggl [ \left( 2\lambda \sqrt{r_1^2-1}\ \right) \left( 2\lambda \sqrt{r_2^2-1}\ \right) \frac{r_1r_2-1-\sqrt{(r_1^2-1)(r_2^2-1)}\cosh(t_1-t_2)}{2\sqrt{(r_1^2-1)(r_2^2-1)}}\biggl] \ .
\end{equation}
Using \eqref{gl2}, and removing the regulator term  $2\lambda \sqrt{(r^2-1)} $, the two point correlator becomes:
\begin{equation}\label{bl2pt}
\langle VV\rangle\sim e^{-m D}= \biggl(\frac{2\sqrt{r_1^2-1} \sqrt{r_2^2-1}}{r_1 r_2 -1-\sqrt{r_1^2-1}\sqrt{r_2^2-1}\cosh(t_1-t_2)}\biggl)^{m} \ .
\end{equation}
%
This matches exactly with the boundary computation as we now show.


\vskip 0.5cm
\underline{\textbf{Boundary computation of the two point function}}:
\vskip 0.5cm
The boundary theory lives at $\phi=0$. From the curve equations in \eqref{a5} in the heating phase, at $\phi=0$ we get:
\begin{align}\label{bdy curve1}
    x+i \tau=z=\sqrt{d}\tan{\mu(n+i \theta)} \ , \\
    x-i \tau=\bar{z}=\sqrt{d}\tan{\mu(n-i \theta)} \ .
\end{align}
If we define $\omega=\mu(n+i \theta)$, then the above equations become:
    \begin{align}\label{bbm1}
        z=\sqrt{d}\tan{\omega} \ , \\
        \label{bbm100}
        \bar{z}=\sqrt{d}\tan{\bar{\omega}} \ .
    \end{align}
Using \eqref{bbm1} and \eqref{bbm100} the two point function can be written as:
\begin{align*}
    \langle \Phi(\omega_{1},\bar{\omega_{1}}) \Phi(\omega_{2},\bar{\omega_{2}})\rangle &=\left(\frac{\partial\omega_{1}}{\partial z_{1}}\right)^{-h}\left(\frac{\partial\omega_{2}}{\partial z_{2}}\right)^{-h}\left(\frac{\partial\bar{\omega_{1}}}{\partial \bar{z_{1}}}\right)^{-h}\left(\frac{\partial\bar{\omega_{2}}}{\partial\bar{ z_{2}}}\right)^{-h}\frac{1}{(z_{1}-z_{2})^{2h}(\bar{z_{1}}-\bar{z_{2}})^{2h}}\nonumber\\
    &= \frac{2^{2h}}{(\cos{2\mu (n_{1}-n_{2})}-\cosh{2\mu (\theta_{1}-\theta_{2})})^{2h}} \ .
\end{align*}
We analytically continue $n\rightarrow in$ to get the Lorentzian correlator as:
\begin{align}\label{bbm3}
    \langle \Phi(\omega_{1},\bar{\omega_{1}}) \Phi(\omega_{2},\bar{\omega_{2}})\rangle = \frac{2^{2h}}{(\cosh{2\mu (n_{1}-n_{2})}-\cosh{2\mu (\theta_{1}-\theta_{2})})^{2h}} \ .
\end{align}
Re-defining $t=2\mu n$ and $\coth{2\mu\theta}=r$,
\begin{equation}\label{bbm2}
    \cosh{2\mu(\theta_{1}-\theta_{2})}=\frac{r_{1}-r_{2}}{\sqrt{(r_{1}^{2}-1)(r_{2}^{2}-1)}} \ .
\end{equation}
Substituting $\eqref{bbm2}$ in $\eqref{bbm3}$, we get:
\begin{align}\label{bd2pt}
    \langle \Phi(\omega_{1},\bar{\omega_{1}}) \Phi(\omega_{2},\bar{\omega_{2}})\rangle =\left[ \frac{2{\sqrt{(r_{1}^{2}-1)(r_{2}^{2}-1)}}}{(\cosh{(t_{1}-t_{2})}\sqrt{(r_{1}^{2}-1)(r_{2}^{2}-1)}-r_{1}r_{2}+1)}\right]^{2h} \ .
\end{align}
For heavy operators identifying $m\sim 2h_v$, we get an exact match with the bulk answer given in \eqref{bl2pt}.

\subsubsection{2-point correlation function in the non-heating phase}


We can rewrite the metric \eqref{me1} corresponding to the non-heating phase ($d>0$) as
\begin{equation}\label{mb1}
    ds^2=\frac{d\phi^2}{\sin^2\phi}+\frac{1}{\sin^2\phi}\biggl(\frac{dr^2}{r^2+1}-(r^2+1)dt^2\biggl) \ .
\end{equation}
by considering the following coordinate change
\begin{equation}\label{cb2}
    r=\cot(2\mu \theta) \ \text{and  } \ t=2\mu s \ .
\end{equation}
The embedding coordinates are given by:
\begin{align}\label{e2}
    T_1 & = \sqrt{r^2+1}\sin t\csc\phi\ , \nonumber\\
    T_2 & =   \sqrt{r^2+1}\cos t\csc\phi \ , \nonumber\\
    X_1 & = \cot\phi  \ , \nonumber \\
    X_2 & =  r \csc\phi \ .
\end{align}
The geodesic length, given by $\eqref{gl1}$, is:
\begin{equation}\label{gl3}
    \cosh D =\lambda^2(\sqrt{(r_1^2-1)(r_2^2-1)}\cosh(t_1-t_2)-1-r_1 r_2) \ .
\end{equation}
Once again, the distance is divergent and we need to regulate it. Hence, as in the heating phase case, we obtain
\begin{equation}\label{gl4}
    D \sim\log\biggl [ \left( 2\lambda \sqrt{r_1^2+1}\ \right) \left( 2\lambda \sqrt{r_2^2+1}\ \right) \frac{\sqrt{(r_1^2+1)(r_2^2+1)}\cosh(t_1-t_2)-r_1r_2-1}{2\sqrt{(r_1^2+1)(r_2^2+1)}}\biggl] \ .
\end{equation}
Using \eqref{gl4}, the two point correlator is obtained to be:
\begin{equation}\label{bl3pt}
    \langle VV\rangle\sim e^{-m D}= \biggl(\frac{2\sqrt{r_1^2+1} \sqrt{r_2^2+1}}{\sqrt{r_1^2+1}\sqrt{r_2^2+1}\cosh(t_1-t_2)-r_1 r_2 -1}\biggl)^{m} \ .
\end{equation}
In the above, the geodesic distance has been regulated with a regulator $2\lambda \sqrt{(r^2+1)} $.

\vskip 0.5cm
\underline{\textbf{The boundary computation}}:
\vskip 0.5cm
At $\phi=0$, the curve equations in \eqref{a3} gives:
\begin{align}\label{bdy curve2}
    x+i \tau=z=-\sqrt{d}\coth{\mu(n+i \theta)}= -\sqrt{d} \coth{\omega} \ , \\
    x-i \tau=\bar{z}=-\sqrt{d}\coth{\mu(n-i \theta)}=-\sqrt{d}\coth{\bar{\omega}} \ .
\end{align}
Using the above equations the two point function in this case can be written as:
\begin{align*}
    \langle \Phi(\omega_{1},\bar{\omega_{1}}) \Phi(\omega_{2},\bar{\omega_{2}})\rangle &=\left(\frac{\partial\omega_{1}}{\partial z_{1}}\right)^{-h}\left(\frac{\partial\omega_{2}}{\partial z_{2}}\right)^{-h}\left(\frac{\partial\bar{\omega_{1}}}{\partial \bar{z_{1}}}\right)^{-h}\left(\frac{\partial\bar{\omega_{2}}}{\partial\bar{ z_{2}}}\right)^{-h}\frac{1}{(z_{1}-z_{2})^{2h}(\bar{z_{1}}-\bar{z_{2}})^{2h}}\nonumber\\
    &= \frac{2^{2h}}{(\cos{2\mu (n_{1}-n_{2})}-\cos{2\mu (\theta_{1}-\theta_{2})})^{2h}} \ .
\end{align*}
The Lorenzian correlator for non-heating case after analytic continuation ($n\rightarrow i n$)  and coordinate re-definition ($t=2\mu n$, $r= \cot{2\mu\theta}$) is:
\begin{align}\label{2pnhbd}
    \langle \Phi(\omega_{1},\bar{\omega_{1}}) \Phi(\omega_{2},\bar{\omega_{2}})\rangle =\left[ \frac{2{\sqrt{(r_{1}^{2}+1)(r_{2}^{2}+1)}}}{(\cosh{(t_{1}-t_{2})}\sqrt{(r_{1}^{2}+1)(r_{2}^{2}+1)}-r_{1}r_{2}-1)}\right]^{2h} \ .
\end{align}
Which matches with the expression derived from the bulk in equation \eqref{bl3pt}

\subsubsection{2-point correlation function in the phase boundary}
The embedding coordinates for this case can be written down as:
\begin{align}\label{e2}
    T_1 & =     \frac{1}{2 r}(1+r^2(1-t^2)) \csc\phi \ , \nonumber\\
    T_2 & =  \cot\phi  \ , \nonumber\\
    X_1 & = t\  r \csc\phi \ , \nonumber \\
    X_2 & = \frac{1}{2 r}(1-r^2(1+t^2)) \csc\phi \ .
\end{align}

In this case the corresponding metric in \eqref{m3}, after the coordinate change $  r=\frac{1}{\theta} \ \text{and  } \ t= s $, can be rewritten as,
\begin{equation}\label{mb3}
    ds^2=\frac{d\phi^2}{\sin^2\phi}+\frac{1}{\sin^2\phi}\biggl(\frac{dr^2}{r^2}-r^2 dt^2\biggl) \ .
\end{equation}

The corresponding geodesic length is :
\begin{equation}\label{gl5}
    \cosh D=\frac{\lambda^2}{2 r_1 r_2}((r_1 -r_2)^2 - (r_1 r_2)^2(t_1-t_2)^2) \ .
\end{equation}
Similar to the previous cases,
%
using \eqref{gl5}, we find the regulated geodesic length, with regulator ($ r \lambda=\frac{\csc\phi}{\theta}$) and then the two point correlator is obtained to be:
\begin{equation}\label{bl4pt}
    \langle VV\rangle\sim e^{-m D}= \biggl(\frac{r_1^2 r_2^2}{(r_1-r_2)^2-r_1^2 r_2^2 (t_1-t_2)^2}\biggl)^{m} \ 
\end{equation}
\vskip 0.5cm
\underline{\textbf{The boundary computation}}:
\vskip 0.5cm
As before, we start with \eqref{a6} at $\phi=0$ and find the two point correlator after analytically continuing $n\rightarrow in$ and suitably redefining coordinates $t= n \ , r= \frac{1}{\theta}$ to be:
\begin{align}\label{2pptbd}
    \langle \Phi(\omega_{1},\bar{\omega_{1}}) \Phi(\omega_{2},\bar{\omega_{2}})\rangle =\left[ \frac{(r_{1}^{2} r_{2}^{2})}{(r_{1}-r_{2})^2-r_{1}^{2} r_{2}^{2}(t_{1}-t_{2})^2}\right]^{2h} \ .
\end{align}
Again this matches exactly with \eqref{bl4pt}.

    \subsection{4-point out of time order correlators from the bulk}

In this section we compute a $4$-point OTOC in the bulk geometry
following the work of  \cite{Roberts:2014ifa1, Shenker:2013pqa1}.
The idea, as  argued in \cite{Shenker:2013pqa1}, is  that the four
point OTOC in the bulk can be thought of as a two point function in
a perturbed shock wave geometry created by one of the operators. In
this section, we will set up the computation in the heating phase
geometry. We will show the emergence of a exponential temporal
behaviour at late times with a Lyapunov exponent which will exactly
match with the boundary value obtained in \cite{Das:2022jrr1}. We
then end the section by pointing out the crucial difference with the
other two phases, which will lead to a non-exponential temporal
behaviour in those cases.

\subsubsection{The shock wave profile}

We begin with a derivation of the shock-wave profile following the seminal work of \cite{Dray:1984ha}. We start with the form of metric given in \ref{ma1}:

\begin{equation}
        ds^2=\frac{d\phi^2}{\sin^2\phi}+\frac{1}{\sin^2\phi}\biggl(\frac{dr^2}{r^2-1}-(r^2-1)dt^2\biggl) \ .
    \end{equation}.

In terms of Kruskal coordinates, this takes the form:
 \begin{equation}\label{ma2}
    ds^2=\frac{d\phi^2}{\sin^2\phi}+\frac{1}{\sin^2\phi}\biggl(\frac{-4}{(1+uv)^2} du dv\biggl) \ ,
 \end{equation}

where, $u=-e^{-\tilde{u}} \ , \ v=e^{\tilde{v}}$ with $\tilde{u}=t-r_* \ ,\ \tilde{v}= t+r_* $ and $r_*=\frac{1}{2}\ln\frac{|r-1|}{r+1}$.

The metric \eqref{ma1} has a horizon at $r=1$ or $u v =-1$. This will then represent a two-sided black-hole geometry in extended Kruskal coordinates. The boundary theory lives at $\phi=0$ hyper-surface.
The above metric \eqref{ma2} is of the following form \cite{Sfetsos:1994xa}:

\begin{equation}\label{ma3}
    ds^2=2 A(u,v) h(\phi) du dv+ h(\phi) d\phi^2 \ ,
\end{equation}
where $A(u,v)=\frac{-4}{(1+uv)^2}$ and $h(\phi)=\frac{1}{\sin^2
\phi}$.  Consider a scenario where a massless particle at $u=0$
moves along the $v$-direction in the background metric \eqref{ma3},
along a constant ($\phi =a$) which back-reacts and results in a
shock wave geometry. Following \cite{Dray:1984ha}, our ansatz for
the form of the shock wave geometry is:
\begin{equation}\label{ma4}
        ds^2=2 A(\tilde{u},\tilde{v}) h(\tilde{\phi}) d\tilde{u} d\tilde{v}-2 A(\tilde{u},\tilde{v}) h(\tilde{\phi}) \eta \delta(\tilde{u}) d\tilde{u}^2 + h(\tilde{\phi}) d\tilde{\phi}^2 \ .
\end{equation}
This shock wave geometry in \eqref{ma4} is described by \eqref{ma3} for both $u>0$ and $u<0$ with the effect of the shock wave being that the $v$ coordinate for $u>0$ is shifted to $v+\eta(\phi)$.
In \eqref{ma4}, $\tilde{v}=v+\eta(\phi)\theta(u)$, $\tilde{u}=u$ and $\tilde{\phi}=\phi$.
Our main objective is to determine $\eta(\phi)$ that determines the shock wave profile \footnote{Let us note that due to the presence of an overall conformal factor $h(\phi)$, the metric in \eqref{ma3} is slightly different from the form of the metric
 considered in \cite{Dray:1984ha} and \cite{Sfetsos:1994xa}. Therefore, we expect the conditions (eg. see Eq. 2.10 of \cite{Sfetsos:1994xa}) on metric components and the equation satisfied by the shock wave profile in our case would be different.}.

The core idea of this calculation is based on the fact that the ansatz metric (\ref{ma4}) solves the Einstein equation with appropriate source terms. These source terms are given by the sum of stress tensor of the unperturbed geometry and the stress tensor of the moving particle ($T^p$) with momentum $p$. Here,
\begin{equation*}
    T^p=T^p_{\tilde{u}\tilde{u}} d{\tilde{u}}^2=- 4 p\ A^2\ h^2(\tilde{\phi})\ \delta(\tilde{u})\ d{\tilde{u}}^2 \ ,
\end{equation*}
Subsequently, comparing the coefficients of $\delta(\tilde{u})$ on both sides of the Einstein equation, we get the following conditions:
 \begin{align*}
    &\text{At } \ \tilde{u}=0\ , \quad  A_{\tilde{v}}=0 \ ,     \\
    & \eta''(\phi)+\frac{h'(\phi)}{h(\phi)}\eta'(\phi) = 32 \pi p\ A\ h^2\ \delta(\phi-a) \ .
 \end{align*}
In our case, this takes the following form:
\begin{equation}
\eta''(\phi)-\cot\phi \eta'(\phi) = -\frac{c' \sin a}{\sin^4\phi}\ \delta(\phi-a) \ ,
\end{equation}
where, $c'= 32\pi\ p$. The solution to above equation is:
\begin{equation}\label{profl0}
    \eta(\phi)=c_2-c_1 \cos\phi + c' \csc^4 a \biggl[(\cos\phi-\cos a)\Theta(\phi-a)\biggl]  \ .
    \end{equation}
    To proceed further, we need to impose boundary conditions to fix the constant $c_1$ and $c_2$. We impose the boundary condition that the shock wave is entirely in the bulk and has no component along the boundary, i.e.: $\eta(\phi)=0$ at $\phi=0,\pi$.

This completely determines the profile function, which takes the following form\footnote{The final expression \eqref{profl} depends quite non-trivially on the specific choice of boundary conditions i.e. $\eta(\phi)=0$ at $\phi=0,\pi$. Given the fact that the calculation's final outcome is heavily reliant on this shock wave profile, one may wonder if there is a particular and distinctive way to select the boundary conditions and whether different boundary conditions will correspond to completely different physical cases. It would be nice to explore these questions further.}.
 \begin{equation}\label{profl}
    \eta(\phi)=\frac{c' \csc^4 a}{2}\biggl[(1-\cos\phi)(1+\cos a)+ 2(\cos\phi-\cos a)\Theta(\phi-a)\biggl]  \ .
    \end{equation}


\subsubsection{4-point OTOC in the heating phase geometry}

Let us now compute OTOC of two scalar operators ($V$ and $W$) in the
bulk. As mentioned earlier, this reduces to the computation of a two
point function in a perturbed shock wave
geometry\cite{Shenker:2013pqa1}.  Therefore, we have to compute
$_{W}\langle V_L V_R\rangle_{W} $ similar to \ref{A} but in the
shock wave geometry produced by a particle $W$, where the two
boundary operators ($V_L,V_R$) with mass $m$ are considered to be on
left and right boundary of the extended geometry. The shock wave in
this case is due to back-reaction produced by the large blue shifted
proper energy, denoted by $E_w$, of the probe particle $W$. This $W$
particle is released from the boundary in the far past, at a time
$t_w$, as measured by a static observer near horizon at time
$t=0$.\footnote{See \ref{a23} for more detail.} We can redefine
$t=2\beta\sqrt{d}s$ and $r'=2\beta\sqrt{d}r$, the metric
\eqref{ma1} becomes a AdS$_{2}$ blackhole patch of AdS$_{3}$ with
horizon at $r'=2\sqrt{d}\beta$ :
\begin{align}
ds^2=\frac{d\phi^2}{\sin^2\phi}+\frac{1}{\sin^2\phi}\biggl(\frac{dr'^2}{r'^2-4d\beta^{2}}-(r'^2-4d\beta^{2})ds^2\biggl).
\end{align}
The boosted large energy at time $t'_{w}=0$ in the above metric is:
\begin{equation}
E_w \sim \frac{E}{4d\beta^{2}} \sin a e^{2\sqrt{d}\beta n_w} \ .
\end{equation}

As we do in the two-point function computations, we start by writing down geodesic lengths in terms of embedding coordinates but this time we write two separate geodesic distances $d_1$ and $d_2$ from a boundary point to some bulk point on both sides of the shockwave geometry. The actual geodesic is then calculated by extremizing the sum of two distances $d_1+d_2$, with respect to $v$ and $\phi$ so that it meets the shock wave at $v_*$ on $\phi_*$ slice. Here, $d_1$ refers to the geodesic length from the left boundary point $(t_L=0, r, \phi_0=0)$ to some bulk point at $(u=0,v,\phi)$, while, $d_2$ is the geodesic length from $(\tilde{u}=u=0,\tilde{v},\phi)$ to the right boundary point $(t_R=0, r, \phi_0=0)$. The expressions for $d_1,d_2$  in terms of embedding coordinates in \eqref{e1} are given by
\begin{eqnarray}\label{d11}
    \cosh d_1= \bigl[r+e^{-t_L}\sqrt{r^2-1}\ v-\cos\phi\cos\phi_0\bigl]\csc \phi\csc\phi_0 \ ,
\\
\label{d21}
\cosh d_2= \bigl[r+e^{-t_R}\sqrt{r^2-1}\ (v+\eta(\phi))-\cos\phi\cos\phi_0\bigl]\csc \phi\csc\phi_0  \ .
    \end{eqnarray}
Recall that for $\phi_0=0$, $\csc\phi_0=\lambda$ diverges and needs a regularization. The final geodesic length is calculated in two steps: First, by extremizing $d_1+d_2$ in \eqref{d11} and \eqref{d21} with respect to $v$ yields: $v_*=-\eta/2$ and the corresponding geodesic length is given by
\begin{equation}\label{e1}
    \cosh \frac{\tilde{d}}{2} = \lambda \left(r+\sqrt{r^2-1}\frac{\eta(\phi)}{2}-\cos\phi \right)\csc\phi \ .
\end{equation}
Extremizing further with respect to $\phi$ yields:
\begin{eqnarray*}
\cos\phi_* = \frac{4+c'\sqrt{r^2-1}(\cos a+1)\csc^4 a}{4r + c'\sqrt{r^2-1}(\cos a+1)\csc^4 a} \ , \quad    \ \ \ \text{for} \quad \ \phi<a \ , \\
\cos\phi_* = \frac{4+c'\sqrt{r^2-1}(\cos a-1)\csc^4 a}{4r + c'\sqrt{r^2-1}(\cos a-1)\csc^4 a} \ .\quad     \ \ \ \text{for} \quad \ \phi>a \ .
\end{eqnarray*}
Substituting $\phi_*$ back in $\eqref{e1}$, the final geodesic distance turns out to be:
\begin{equation}
d\approx2 \log \left[2 \lambda\sqrt{r^2-1} \right] +\log \left[1+ c' \sqrt{\frac{r-1}{r+1}}\biggl(\frac{\cos{a}+1}{\sin^4{a}}\biggl) \right] \ \ \ , \ \text{for } a>\frac{\pi}{2} \ ,
\end{equation}
or,
\begin{equation}
    d\approx2\log \left[2 \lambda\sqrt{r^2-1} \right] +\log \left[1+ c' \sqrt{\frac{r+1}{r-1}}\biggl(\frac{1-\cos{a}}{\sin^4{a}}\biggl) \right] \ \ \ , \ \text{for } a<\frac{\pi}{2} \ .
\end{equation}

After subtracting the divergent contribution from $2\lambda\sqrt{r^2-1}$ and using a geodesic approximation, $_{W}\langle VV \rangle_{W}\propto e^{-m d}$ with regularized geodesic distance $d$ and substituting $c'\sim 32\pi \frac{E}{4d\beta^{2}}\sin a e^{2\sqrt{d}\beta n_w}$, we find that the final form of OTOC is:
\begin{equation}\label{otoc}
    \frac{_{W}\langle V_L V_R \rangle_{W}}{\langle W W\rangle\langle V_L V_R\rangle} \approx \left(\frac{1}{1+ 32\pi \frac{E}{4d\beta^{2}}  \sqrt{\frac{r\mp1}{r\pm1}}\bigl(\frac{1\pm\cos{a}}{\sin^3{a}}\bigl)e^{2\sqrt{d}\beta n_w}} \right)^m \ ,
\end{equation}
From the above expression we get the Lyapunov exponent to be $2
\sqrt{d} \beta$ . This matches precisely with the Lyapunov exponent
obtained from a purely CFT computation in \cite{Das:2022jrr1}. In
that work, a direct and explicit CFT calculation was carried out for
a large $c$ CFT.  With a discrete drive, in the heating phase, the
four point OTOC in large-c CFT was obtained to be:
\begin{equation}\label{fpbo}
    \mathcal{F}=\left(\frac{1}{1-\frac{24\pi ih_{\text{w}}e^{4n\theta}}{c \epsilon_{12}\epsilon_{34}A(z_{\text{w}},z_{\text{v}})}}  \right)^{2h_{\text{v}}} \ ,
\end{equation}
where, $A(z_{\text{w}},z_{\text{v}})=-16\theta^2\frac{(z_{\text{v}}-1)(z_{\text{w}}+1)}{(z_{\text{v}}+1)(z_{\text{w}}-1)}$. From the above equation, one can extract the Lyapunov exponent to be:
\begin{align}\label{fpbo2}
    \lambda_{\text{L}}=\frac{4\theta}{(T_{1}+T_{2})} \ .
\end{align}
Expressing the above equation in terms of the parameters of the effective hamiltonian, then using \eqref{eff101}, we get:
\begin{align}\label{fpbo3}
    \lambda_{\text{L}}&=\sqrt{\alpha^2-4\beta\gamma} = 2\beta\sqrt{d} = \frac{4\theta}{(T_{1}+T_{2})}  \ .
\end{align}
This matches with the bulk computation. It would be nice to match
the full expression \eqref{fpbo} with its bulk counterpart and not
just the Lyapunov exponent. The function $A(z_{\text{v}},
z_{\text{w}}$) is a non-trivial function of the position of two
operators $V$ and $W$.  The bulk expression derived here is a
function of the position of the boundary $V$ operator, however the
only information of the $W$ operator which enters is the direction
$\phi =a$ along which the particle which creates the shock wave
propagates. We have not been able to translate this information into
the boundary location of the $W$ operator. However, it is
encouraging that the dependence on the position of the $V$ operator
is similar in both the expressions. We hope to be able to return to
this in the near future.


\subsubsection{OTOC in non-heating phase and phase transition}\label{a23}

Let us repeat the same calculations in the non-heating phase, as well as on the phase boundary. In the heating case, the exponential behaviour in the OTOC was due to the shockwave geometry that results from the large blue-shifted energy $\cO(e^t_w)$ of the $W$ particle which is released at a very early time $t_w$. In general, if a particle released from the boundary $r\rightarrow\infty$ at an early time is moving along a null trajectory with proper energy $E$, the energy $E_r$ measured on the time slice $t=0$, is
\begin{equation}\label{pe}
    E_r=\frac{E}{\sqrt{g_{00}|_{t=0}}} \ .
\end{equation}
We will now investigate the behavior of $E_r$ for the metrics in other phases.
\vskip 0.2cm
\textbf{\underline{For non-heating phase}}:
We start with the metric \eqref{me1} and follow exact similar procedure as in the previous section to rewrite the metric in terms of $r=\cot(2\mu\theta)$, $t=2\mu s$ and $\phi$:
\begin{equation}\label{mb1}
    ds^2=\frac{d\phi^2}{\sin^2\phi}+\frac{1}{\sin^2\phi}\biggl(\frac{dr^2}{r^2+1}-(r^2+1)dt^2\biggl) \ .
\end{equation}
The tortoise coordinate $r_*$ in this case is given by $\frac{dr_*}{dr}=\frac{1}{1+r^2}$ and hence, $r_*=\tan^{-1}r$. Therefore, the trajectory of nearly null $W$-particle released from boundary at $t_w$, in terms of the tortoise coordinate $r_*$ at time t is:
\begin{equation}\label{traj}
    t-t_w=r_*-\frac{\pi}{2} \ .
    \end{equation}
Substituting \eqref{traj} in \eqref{pe} we see that for metric \eqref{mb1} the energy measured at time $t=0$ is:
\begin{equation}\label{enh}
    E_r= E \sin{a}\ \sin{t_w} \ .
\end{equation}
 \vskip 0.1cm
 \textbf{\underline{On the phase boundary}:}\\
Similarly, the metric \eqref{m3} can be re-defined in terms of $r=\frac{1}{\theta}$ and $n=t$. The tortoise coordinate for this case is given by $r_*=\int\frac{dr}{r^2}=-\frac{1}{r} $. Then the null trajectory of $W$-particle in this case is:
\begin{align*}
    \int^t_{t_w} dt= \int^{r_*}_0 dr\ , \\
\Rightarrow t-t_w=r_* \ .
\end{align*}
Then using this in \eqref{pe} we find the energy measured at $t=0$ is:
\begin{equation}\label{etl}
    E_r=E \sin a\ t_w \ .
\end{equation}
Hence we find that the energy measured at time $t=0$ in non-heating
phase \eqref{enh} and during phase transition \eqref{etl}
respectively show oscillatory and power law dependence on $t_w$. This
is consistent with the boundary results \cite{Das:2022jrr1}.

\numberwithin{equation}{section}
    \section{Discussions}
    \label{diss}

In this article, we constructed a Holographic description of a
(gravity + brane) system which is capable of detecting the
non-heating to heating phase transition in the dual boundary CFT,
which is subject to a periodic drive. While this framework is
completely natural and intuitive in this respect, our construction
should be viewed as the simplest of the richer possibilities.
\footnote{Note that, the richness of a boundary degree of freedom in
dynamical context has been explored also in \cite{Banerjee:2018twd,
Banerjee:2018kwy, Banerjee:2019vff} in the probe limit and in
\cite{Das:2019tga, Das:2021qsd, Aramthottil:2021cov} away from any
probe approximation.} Subsequently, there are several intriguing
aspects for future explorations. We enlist some of them below.

First, note that the periodically driven Hamiltonian is
$sl(2,R)$-valued and therefore does not accommodate the
possibilities of a large gauge transformation. The general class of
Brown-Henneaux diffeomorphisms contain an infinite number of such
large gauge transformations, which are dual to a periodically driven
Hamiltonian valued in the $sl^{(q)}(2,R)$, for $q>1$. Conceptually,
it is no harder to find the corresponding curves in the bulk which
would be generated by the bulk Hamiltonian. Subsequently, the
various patches will likely contain a richer class of metrics,
including dynamical ones. It will be an interesting question to
consider these cases, in the presence of EOW-branes.

A much simpler problem is to consider the $sl(2,R)$-valued drive
Hamiltonian and work out the corresponding phase patches starting
from a global AdS$_3$. In this case, the CFT is defined on a
cylinder and the corresponding (gravity + brane) on-shell action
corresponds to the boundary entropy of the dual BCFT. This boundary
entropy counts the ground-state degeneracy in the BCFT and it will
be interesting to understand in detail how this counting detects the
phase transition. This further generalizes in the presence of more
than one EOW-branes, with different tensions.

Relatedly, we can explore an alternative way of inserting the
EOW-branes in the bulk. One can begin with a bulk Hamiltonian, in a
geometry where EOW-branes are already inserted and subsequently
analyze the tangent curves and the corresponding induced geometries.
This is conceptually different from what we have done here. Although
we expect the qualitative features to remain the same, especially so
since the EOW-branes emerge naturally in the corresponding patches
that we have considered here, it will nonetheless be an interesting
issue to understand in greater detail.

A crucial point of our study is the appearance of AdS$_{2}$ slicing
which plays a pivotal role in distinguishing phases in terms of
unequal time correlators as well as provides a natural setting to
incorporate EOW brane.  In particular, the OTOC computation strongly
suggests that the AdS$_{2}$ physics is responsible for the
different temporal growth in different phases. From the boundary perspective it is not at
all clear why such AdS$_{2}$ foliation emerges. For instance, from
Eq. (\ref{bdy curve1}) and (\ref{bdy curve2}) the boundary tangent
curve parametrizes a time dependent boundary metric. This time
dependence in boundary metric is responsible for the different
temporal behavior of the unequal time correlators. However when we
lift those boundary metrics to the bulk AdS$_{3}$ we end up with
time independent AdS$_{2}$ slicing of AdS$_{3}$. 
The presence of
the EOW-branes in an AdS-background suggests a doubly-Holographic
model structure. Such models have recently been intensely explored
in connection with the black hole information paradox
\cite{Penington:2019npb}-\cite{Rozali:2019day}. It will be
intriguing if there is a clear connection between the physics of the
transition with the physics of the information paradox here. We hope
to return with a more clear answer in future.

Relatedly, it will be interesting to construct examples in which the
black hole on the brane becomes truly dynamical. This aspect is
expected to be visible with a periodic drive with an
$sl^{(q)}(2,R)$-valued Hamiltonian. Alternatively, similar dynamical
situation could be appeared in a primary state under the $sl(2,R)$
drive. We would like to address some of these issues in future.

\section{Acknowledgements}

We would like to thank Tarek Anous, Parthajit Biswas, Pawel Caputa, Diptarka Das, Damian Galante, Dongsheng Ge, Chethan Krishnan, Sinong Liu, Vinay Malvimat, Giuseppe Policastro, Koushik Ray and Ritam Sinha for useful conversations on related topics. SD would like to acknowledge the support provided by the Max Planck Partner Group grant MAXPLA/PHY/2018577. BE is supported by
CRG/2021/004539, AK acknowledges support from the Department of
Atomic Energy, Govt. of India, Board of Research in Nuclear Sciences
(58/14/12/2021-BRNS) and IFCPAR/CEFIPRA 6304-3. The work of SP and
BR is supported by a Senior Research Fellowship(SRF) from UGC. KS
thanks DST, India for support through project JCB/2021/000030. We
would also like to thank the Organizers of Out of Equilibrium
Physics, held at Indian Institute of Technology, Mandi for a
stimulating environment where preliminary results related to this
project were presented. SD, SP and BR would like to thank the organizers of ST$^{4}$ 2022, held at Indian Institute of Technology, Indore, where part of the work have been discussed. SD would like to acknowledge the hospitality of University of Amsterdam and University of Warsaw where part of the work was presented. BE would like to acknowledge the hospitality of the Physics department at IIT Kanpur, where some of the results were presented.

    \appendix
    \numberwithin{equation}{section}
\section{Floquet (Effective) Hamiltonian of a driven CFT}\label{B}

\begin{figure}
    \centering
    \includegraphics[width=0.50\linewidth]{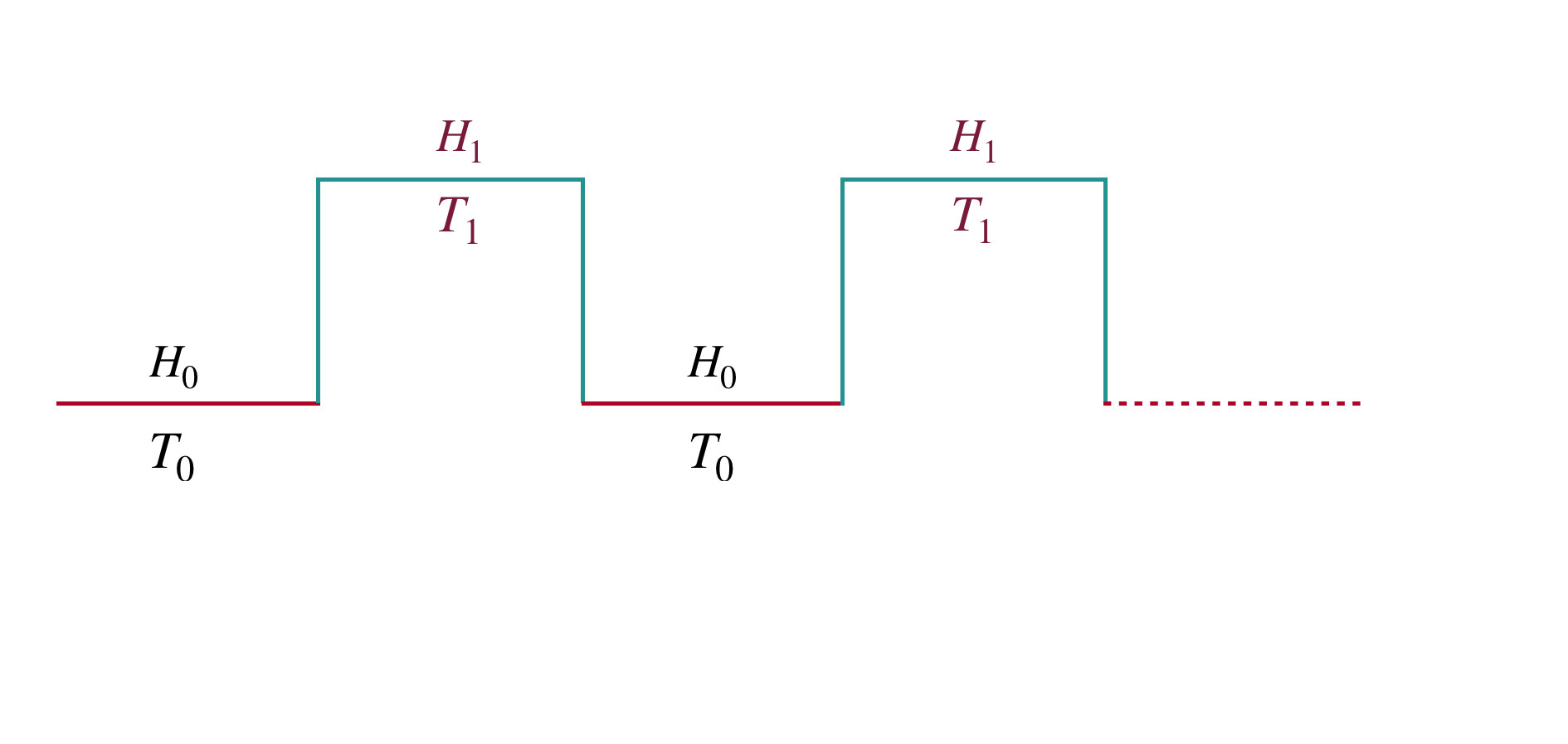}
    \caption{Pictorial representation of discrete drive protocol}
    \label{fig:drive}
\end{figure}

In this appendix, we explicitly construct the Floquet (Effective) Hamiltonian ($H_{eff}$) for the two period discrete drive protocol.
 The form of Floquet Hamiltonian depends on the driving protocol of the CFT. As an example, we compute the Floquet Hamiltonian of a discretely (two-step) driven CFT where, where, the Hamiltonian $H_{\theta}=\int_{0}^{L}T_{00}(1-\tanh{(2\theta)}\cos(\frac{2\pi x}{L}))dx$ in each period switches between $H_{0}=H_{\theta=0}\ , H_{1}=H_{\theta\neq 0}$
as in Fig.\ \ref{fig:drive} \cite{Wen:2020wee}.

 In terms of the modes, the Hamiltonians are given as following:\
\begin{equation*}
    H_{0}=\frac{2\pi}{L}\left[L_{0}+\bar{L_{0}}\right]-\frac{\pi c}{12L} \ ,
\end{equation*} and \vskip 0.1cm
\begin{equation*}
    H_{1}=\frac{2\pi}{L}\biggl[L_{0}-\tanh{(2\theta)}\frac{L_{1}+L_{-1}}{2}\biggl]-\frac{\pi c}{12L}+\text{anti-holomorphic part.}
\end{equation*}
At this point, we make an ansatz for the Floquet Hamiltonian, that replicates the same dynamics as the original system. We assume the following as the Floquet Hamiltonian\footnote{Here we ignore the c-number part which is irrelevant for our purpose.}
\begin{equation*}
    H_{eff}= [\alpha L_{0}+ \beta L_{1}+ \gamma L_{-1}]+\text{anti-holomorphic part(A.H)} \ . 
\end{equation*}
Since $H_{0}$ and $H_{1}$ are only made of $L_{0},L_{\pm 1}$, the BCH formula guarantees that the Floquet Hamiltonian should be made of only by those global conformal generators.
We may able to determine $\alpha,\beta,\gamma$ by demanding that $H_{eff}$ must satisfy the following condition:
\begin{equation}\label{eff1}
    e^{-\tau_{0}H_{0}}e^{-\tau_{1}H_{1}}.z=e^{-(\tau_{0}+\tau_{1})H_{eff}}.z \ . 
\end{equation}
For discrete drive, the LHS of the above equation gives \cite{Wen:2018vux},
\begin{equation}\label{eff2}
    e^{-\tau_{0}H_{0}}e^{-\tau_{1}H_{1}}.z=\frac{[(1-\delta)\cosh{2\theta} -(\delta+1)](\frac{\delta'}{2\sqrt{\delta\delta'}})z+\frac{(\delta-1)}{\sqrt{2 \delta\delta'}} \sinh {2\theta}}{[(1-\delta) \sinh{2\theta}](\frac{\delta'}{2\sqrt{\delta\delta'}}) z+\frac{1}{\sqrt{2 \delta\delta'}}[(\delta-1)\cosh{2\theta}-(\delta+1)]} \ , 
\end{equation}
where $\delta:=e^{\frac{2\pi\tau_{1}}{L\cosh{2\theta}}}$ and $\delta':=e^{\frac{2 \pi\tau_0}{L}}$. First, we'll compute the action of $H_{eff}$ on $z$. We will then compare the result with \eqref{eff2} to derive the relations between the parameters $\alpha, \beta, \gamma$ and the parameters of the drive. In the $z$ plane, $H_{eff}$ is
\begin{equation}\label{eff3}
    H_{eff}=\int \frac{dz}{2\pi i}(\alpha z+\beta z^2+\gamma)T(z)+ \text{A.H} \ .
\end{equation}
To simplify \eqref{eff3} further, we map it to $\tilde{z}$ plane such that $H_{eff}$ becomes
\begin{equation}
    H_{eff}=\int \frac{d\tilde{z}}{2\pi i}\tilde{z}T(\tilde{z})+ {\rm A.H}  =\tilde{L_{0}} \ , 
\end{equation}
where $\tilde{z}$ and $z$ are related by the following transformation:
\begin{equation}\label{eff4}
    \tilde{z}=\left[\frac{c'(z-A)}{z-B}\right]^{\frac{1}{\sqrt{(\alpha^2-4\beta\gamma)}}} \ , 
\end{equation}
where, $c'$ is a constant, $A=\frac{-\alpha+\sqrt{(\alpha^2-4\beta\gamma)}}{2\beta}$ and $B=\frac{-\alpha-\sqrt{(\alpha^2-4\beta\gamma)}}{2\beta}$.
In the $\tilde{z}$ plane, $H_{eff}$ acts as,
\begin{equation}
    e^{-s H_{eff}}\tilde{z}=e^{-s\tilde{L_{0}}}=e^s \tilde{z} \ , 
\end{equation}
using the above identity and \eqref{eff4}, we found that
\begin{align}\label{eff5}
    z'& = e^{-\tau_{0}H_{0}}e^{-\tau_{1}H_{1}}.z =e^{-(\tau_{0}+\tau_{1})H_{\text{eff}}}.z(\tilde{z}) = z(e^{-(\tau_{0}+\tau_{1})}\tilde{z}) \nonumber\\
    &  =\frac{\frac{Am^{\frac{-1}{2}}-Bm^{\frac{1}{2}}}{A-B}z+\frac{AB}{A-B}(m^{\frac{1}{2}}-m^{\frac{-1}{2}})}{(m^{\frac{-1}{2}}-m^{\frac{1}{2}})\frac{z}{A-B}+{\frac{Am^{\frac{-1}{2}}-Bm^{\frac{1}{2}}}{A-B}}} \ . 
\end{align}
here, $m=e^{(\tau_{0}+\tau_{1})\sqrt{(\alpha^2-4\beta\gamma)}}$. After comparing \eqref{eff5} with \eqref{eff2} we found that,
\begin{align}\label{beta,alpha}
    \frac{\beta}{\alpha} & =\frac{\delta^{'}(1-\delta)\sinh{2\theta}}{(\delta+1)(1-\delta^{'})-(\delta-1)(1+\delta^{'})\cosh{2\theta}} \ , \\
    \frac{\gamma}{\alpha}& =\frac{(1-\delta)\sinh{2\theta}}{(\delta+1)(1-\delta^{'})-(\delta-1)(1+\delta^{'})\cosh{2\theta}} \ , \\
    m & =[ \frac{(((1+\delta)(1+\delta^{'})+(1-\delta)(1-\delta^{'})\cosh{2\theta})^{2}-16\delta\delta^{'})^{\frac{1}{2}}}{4(\delta\delta')^{\frac{1}{2}}}+\nonumber \\ &\ \hskip 3cm  \frac{(1+\delta)(1+\delta^{'})+(1-\delta)(1-\delta^{'})\cosh{2\theta}}{4(\delta\delta')^{\frac{1}{2}}}]^{2} \ . 
    \end{align}

In the discrete drive protocol, a simple choice of drive frequencies leads to a heating phase: $i\tau_{0}\equiv T_{0} = \frac{L}{2}, i\tau_{1}\equiv T_{1} =\frac{L\cosh(2\theta)}{2}$ such that the su(1,1) transfer matrix takes a simple form: $a_{n}=d_{n}=(-1)^{n}\cosh(2n\theta); b_{n}=c_{n}=-(-1)^{n}\sinh(2n\theta)$\cite{Wen:2020wee}. We have used this protocol to determine OTOC in discrete drive protocol in \cite{Das:2022jrr1}. Plugging this choice of Lorentzian time periods in \ref{beta,alpha} we get
\begin{align}    \label{eff101}
&\sqrt{\alpha^2-4\beta\gamma} =\frac{4\theta}{(T_{0}+T_{1})}  \ , \\
&\alpha = 0, \; \beta=-\gamma=\frac{2\theta}{T_{0}+T_{1}}  \ . \nonumber
    \end{align}
 Note that, the Floquet Hamiltonian for this choice reduces to\footnote{One may wonder how we get $\alpha^{2}-4\beta\gamma > 0$ in heating phase. This is due to the fact, when we write $U_{\text{eff}}=e^{-(\tau_{0}+\tau_{1})H_{\text{eff}}}$, upon analytic continuation to Lorentzian time, we can write the evolution operator with $H_{\text{eff}}=i\alpha L_{0}+i\beta L_{1} +i\gamma L_{-1}$. This shift of $\alpha,\beta,\gamma \rightarrow i\alpha,i\beta,i\gamma$ changes the sign of the Casimir and hence the sign of $\alpha^{2}-4\beta\gamma$.}
 \begin{align}
 e^{-i(\tau_{0}+\tau_{1})H_{\text{eff}}} = e^{2\theta\left(L_{1}-L_{-1}+\bar{L}_{1}-\bar{L}_{-1}\right)} \ . 
 \end{align}
Interestingly, this Hamiltonian also annihilates the boundary state $|B\rangle$\footnote{By definition, $(L_{n}-\bar{L}_{-n})|B\rangle = 0$.} apart from the vacuum. Thus this Hamiltonian can not distinguish between vacuum and boundary state.

\end{document}